\documentclass[prd,twocolumn,nofootinbib,showpacs,10pt]{revtex4-1}
\usepackage{textcomp}
\usepackage{amsmath}
\usepackage{amsfonts}
\usepackage[bbgreekl]{mathbbol}
\usepackage{calc}
\usepackage{mathrsfs}
\usepackage{amssymb}
\usepackage{amsmath}
\usepackage{array}
\usepackage{color}
\usepackage{bbm}
\usepackage{graphicx}
\usepackage{hyperref}

\renewcommand{\P}{\mathcal{P}}
\newcommand{\res}{\mathcal{R}}

\newcommand{\e}{\epsilon}

\newcommand{\nhat}{{\hat n}}

\newcommand{\GW}[1]{\breve{#1}}

\renewcommand{\O}{\mathcal{O}}

\newcommand{\beq}{\begin{equation}}
\newcommand{\eeq}{\end{equation}}

\begin{document}
\title{Second-order perturbation theory: problems on large scales} 
\author{Adam Pound} 
\affiliation{Mathematical Sciences, University of Southampton, Southampton,
United Kingdom, SO17 1BJ}
%\pacs{04.20.-q, 04.25.-g, 04.25.Nx, 04.30.Db}
\date{\today}

\begin{abstract}
In general-relativistic perturbation theory, a point mass accelerates away from geodesic motion due to its gravitational self-force. Because the self-force is small, one can often approximate the motion as geodesic. However, it is well known that self-force effects accumulate over time, making the geodesic approximation fail on long timescales. It is less well known that this failure at large times translates to a failure at large distances as well. At second perturbative order, two large-distance pathologies arise: spurious secular growth and infrared-divergent retarded integrals. Both stand in the way of practical computations of second-order self-force effects. 

Utilizing a simple flat-space scalar toy model, I develop methods to overcome these obstacles. The secular growth is tamed with a multiscale expansion that captures the system's slow evolution. The divergent integrals are eliminated by matching to the correct retarded solution at large distances. I also show how to extract conservative self-force effects by taking local-in-time ``snapshots'' of the global solution. These methods are readily adaptable to the physically relevant case of a point mass orbiting a black hole. %They should also apply to other systems with slowly evolving frequency spectrums.

\end{abstract}

\maketitle

\section{Introduction}
Compact binary inspirals are perhaps the single most important class of sources for gravitational-wave detectors. Within the next few years, binaries of comparable-mass objects are expected to provide the first direct detection of gravitational waves. In the future, extreme-mass-ratio inspirals (EMRIs), in which stellar-mass compact objects spiral into supermassive black holes, are expected to be observed by a planned space-based detector and to offer peerlessly precise maps of the black holes' spacetime geometry~\cite{eLISA:13, Amaro-Seoane-etal:14,Babak-etal:15}. 

Over the last decade, modeling of compact binaries has reached maturity. For comparable-mass binaries, post-Newtonian (PN) theory~\cite{Blanchet:14,Poisson-Will:14} can be used for most of the inspiral, while the objects are widely separated, and numerical relativity can be used for the late stages. These two methods can then be artfully combined using effective-one-body theory (EOB)~\cite{Buonanno-Damour:99,Damour-etal:00}, which (when appropriately calibrated) provides a fast and accurate method of generating waveform templates. For EMRIs, one can use self-force theory~\cite{Poisson-Pound-Vega:11,Pound:15a}, which provides a model based on an expansion of the Einstein equations in powers of the small object's mass $m$. 

In addition to the development of these distinct models, the last decade has seen their confluence. Comparisons between them have provided essential tests of their validity, and have thus far shown remarkable agreement between them. From the perspective of self-force theory, these successful comparisons have afforded an opportunity to use self-force computations to assist in modeling intermediate-mass-ratio inspirals (IMRIs) and even comparable-mass binaries~\cite{LeTiec-etal:11,LeTiec-etal:12b,LeTiec:14,Bernuzzi-etal:14,LeTiec:15}. This can be done in two ways: directly applying self-force theory to these binaries, a prospect inspired by the surprising accuracy of self-force results outside their expected domain of validity~\cite{LeTiec-etal:11,LeTiec-etal:12b,LeTiec:15}; or alternatively, using self-force data to improve other models by determining high-order PN coefficients and calibrating EOB. So far, the latter approach is the one that has been taken. 

Whether one wishes to model EMRIs, IMRIs, or comparable-mass binaries with self-force theory, one must carry the method to at least second order in $m$ to obtain a sufficiently accurate approximation for waveform modeling~\cite{Hinderer-Flanagan:08,LeTiec:15}. At first order in $m$, numerical implementations are thoroughly developed~\cite{Barack:09,Wardell:15}, can achieve incredible accuracy~\cite{Shah-Pound:15,Johnson-McDaniel-etal:15,Kavanagh-etal:15,Hopper-etal:15}, are now being performed for both Schwarzschild and Kerr backgrounds~\cite{Isoyama-etal:14,vandeMeent-Shah:15}, and have yielded an ever-growing catalogue of computed self-force effects~\cite{Nolan-etal:15}. At second order in $m$, the picture is less rosy. Second-order self-force theory is well developed~\cite{Rosenthal:06a,Rosenthal:06b,Detweiler:12, Pound:12a, Gralla:12, Pound:12b, Pound:15b}, and at the level of formalism all the necessary ingredients are in place~\cite{Pound:12a,Gralla:12,Pound:12b}. But although work is underway to implement this formalism~\cite{Pound-Miller:14,Warburton-Wardell:14,Wardell-Warburton:15}, currently no concrete numerical computations have been performed. 

The primary obstacle to practical implementation is what I will call \emph{the problem of large scales}. Consider an EMRI. At first order in self-force theory, the small object can be treated as a point mass. On the orbital timescale, the particle's worldline deviates from a background geodesic by only a small amount, of order $m$. Hence, if one is interested in effects that occur on this timescale, one can neglect the deviation, approximate the orbit as a geodesic, and solve the linearized Einstein equation with a geodesic source orbit. This comes with an enormous advantage: the source has a discrete frequency spectrum, allowing one to conveniently solve the field equation in the frequency domain.  

However, if we wish to model the long-term inspiral of an EMRI, this approach obviously fails. The deviation from any given geodesic grows large in both the future and the past. Since the deviation acts as a second-order source term in the Einstein equation, this failure manifests as a secularly growing second-order perturbation. Furthermore, because errors propagate at finite speeds, the secular errors at large past times produce growing errors at large distances at fixed time. I call these two types of errors, collectively, the problem of spurious secular growth.

One way to overcome this problem, without giving up the advantages of the geodesic approximation, is to perform a sequence of simulations, each accurate on the orbital timescale, and evolve smoothly between them using a multiscale expansion, osculating-geodesic approximation, or similar scheme~\cite{Mino:06,Hinderer-Flanagan:08,Pound-Poisson:08b,Warburton-etal:12,Pound:15a}. No such scheme has been practically formulated through second order, but an appropriate formulation will allow one to continue to work in the frequency domain. Moreover, even in the absence of such a scheme, without the prospect of evolving between the small-scale solutions, one is often interested in effects that occur on orbital scales. For example, to compare with PN results and calibrate EOB, one can define and compute nondissipative effects at second order that do not require information about the long-term changes of the orbit~\cite{Pound:14c}. %These conservative effects, which do not involve the part of the second-order perturbation sourced by the growing, dissipative shift in the orbit, are again generated by a source with a discrete frequency spectrum. Hence, they can seemingly be computed by solving . 

However, even for calculations on the orbital timescale, a second problem of large scales arises. A bound periodic source is incompatible with asymptotic flatness in general relativity, and in perturbation theory this manifests in an everywhere-divergent retarded solution at second order if we use a periodic source at first order. Mathematically, this divergence occurs because the first-order perturbation generates curvature that acts as a noncompact source for the second-order perturbation. If the first-order solution is periodic, then this second-order source falls off too slowly to be integrated, giving rise to a divergent integral. I call this the  problem of infrared divergence.

In principle, both problems are easily overcome: one can simply solve the coupled system of field equations and equation of motion self-consistently in the time domain, using the self-accelerated orbit as the source, avoiding ever involving a geodesic source. Such a scheme, developed to all perturbative orders in Refs.~\cite{Pound:10a,Pound:12b,Pound:15a} and numerically implemented in a scalar model in Ref.~\cite{Diener-etal:12}, would trivially eliminate the spurious secular growth. And if one begins with asymptotically flat initial data, then the solution will propagate forward with no problems coming from large distances. Unfortunately, such an approach is unlikely to be sufficiently fast or accurate to simulate complete inspirals. An inspiral lasts $\sim 10^5$ orbits in the case of an EMRI, requiring a lengthy run time, and at each small step of this time, the self-force would have to be calculated with very high accuracy to avoid accumulating numerical error in the orbital evolution. Additionally, a self-consistent time-domain evolution would not provide a clean separation into conservative and dissipative effects, since the two combine in highly complicated, nonlinear ways during the evolution.%; as in numerical relativity, separating the full solution into constituent conservative and dissipative parts would be very difficult.%This would likely come with its own difficul

%\cite{Barack-Golbourn:07, Vega-Detweiler:07,Vega-Wardell-Diener:11,Dolan-Barack:13} \cite{Pound:12a,Pound:12b,Pound:15a}

In this paper, I take the self-consistent formulation as my starting point, but from it I seek to derive a more practical, frequency domain, multiscale approach. Following a longstanding tradition in self-force theory~\cite{Barack-Ori:02,Detweiler-etal:03,Haas:07,Barack-Golbourn:07, Vega-Detweiler:07,Diener-etal:12,Warburton:14,Wardell-etal:14,Rosenthal:05,Harte:08,Galley:12a}, to develop my approach I work with a simple scalar toy model. The model, which I specialize to flat spacetime, includes nonlinearities that closely mimic those in gravity, and the tools I devise for it should carry over almost immediately to the gravity case.

I begin in Sec.~\ref{model} by introducing the model. In Sec.~\ref{near-zone}, I then show how the two problems of large scales arise. To eliminate the spurious secular growth, in Sec.~\ref{multiscale} I devise a suitable multiscale expansion. This expansion follows in the tradition of earlier work~\cite{Hinderer-Flanagan:08,Mino-Price:08,Pound-Poisson:08b,Pound:10b,Pound:10c}, particularly the ideas of Hinderer and Flanagan~\cite{Hinderer-Flanagan:08}, but it represents the first instance of a systematic multiscale expansion of a coupled system comprising a nonlinear field equation and an equation of motion. 

To cure the problem of infrared divergences, in Sec.~\ref{matching} I match the multiscale expansion to a general expansion of the exact retarded solution at very large distances. That retarded solution is constructed using the post-Minkowski (PM) methods developed by Blanchet, Damour, and Iyer~\cite{Blanchet-Damour:86,Blanchet:87,Blanchet-Damour:88, Blanchet-Damour:89, Damour-Iyer:91b, Blanchet-Damour:92, Blanchet:14} and by Will and collaborators~\cite{Wiseman-Will:91, Will-Wiseman:96, Pati-Will:00,Pati-Will:01,Poisson-Will:14}. Matching to that solution makes my approach very similar in spirit to PN theory, where a near-zone PN expansion is matched to a far-zone PM one; but in my case, the fully relativistic multiscale expansion replaces the PN expansion. The notion of matching a multiscale expansion to a PM expansion, like the multiscale expansion itself, was first suggested by Hinderer and Flanagan~\cite{Hinderer-Flanagan:08}. %For particular PM tools, my analysis will rely specifically on Refs.~\cite{Blanchet-Damour:86,Blanchet-Damour:88}, which I will refer to as BD-I and BD-II, and Ref.~\cite{Poisson-Will:14}, which I will refer to as PW.

Combining multiscale and PM methods in this way leads to a globally accurate solution. However, if one is interested only in local effects, such as the one described in Ref.~\cite{Pound:14c}, then instead of a full multiscale expansion, one can construct ``snapshot'' solutions on short timescales, again avoiding the infrared divergences by matching to the correct retarded solution at large distances. This construction is set out in Sec.~\ref{snapshots}.

Finally, in Sec.~\ref{lessons} I describe how these methods apply to the gravity case, and I discuss the remaining obstacles to performing a second-order self-force  computation.

A recurring requirement of the analysis will be a change from one time variable to another. To avoid ambiguity, for any function $f:x\mapsto f(x)$, I will always avoid the cavalier notation $f(x(y))=f(y)$; instead, I will always introduce a new function, as in $F(y)=f(x(y))$. An overdot, as in $\dot f$, will represent a derivative with respect to the argument (or to the first argument, in the case of multiple arguments).

%%%%%%%%%%%%%%%%%%%%%%%%%%%%%%%%%%%%%%%%%%%%%%%%%%%%%%%
\section{The toy model}\label{model}

\subsection{Second-order gravity}
To motivate my choice of toy model, I first summarize the second-order gravitational self-force approximation.

In the self-consistent scheme~\cite{Pound:10a}, the Einstein equations in the Lorenz gauge are~\cite{Pound:12a,Pound:12b}
\begin{align}
E_{\mu\nu}[h^1] &= -16\pi \bar T_{\mu\nu}(x;z),\label{EFE1}\\
E_{\mu\nu}[h^{2\res}] &= 2\delta^2 R_{\mu\nu}[h^1]-E_{\mu\nu}[h^{2\P}].\label{EFE2}
\end{align}
Here $E_{\mu\nu}[h]:=\Box h_{\mu\nu}+2R_\mu{}^\alpha{}_\nu{}^\beta h_{\alpha\beta}$ is the Lorenz-gauge wave-operator of the background metric $g_{\mu\nu}$, $T_{\mu\nu}(x;z)$ is the stress-energy at point $x$ of a point mass moving on a worldline $z^\mu$, $\bar T_{\mu\nu}$ is its trace reverse, and
\begin{align}
\delta^2R_{\alpha\beta}[h] &:=-\tfrac{1}{2}\bar h^{\mu\nu}{}_{;\nu}\left(2h_{\mu(\alpha;\beta)}-h_{\alpha\beta;\mu}\right) 
					+\tfrac{1}{4}h^{\mu\nu}{}_{;\alpha}h_{\mu\nu;\beta}\nonumber\\
					&\quad  +\tfrac{1}{2}h^{\mu}{}_{\beta}{}^{;\nu}\left(h_{\mu\alpha;\nu} - h_{\nu\alpha;\mu}\right)\nonumber\\
					&\quad-\tfrac{1}{2}h^{\mu\nu}\left(2h_{\mu(\alpha;\beta)\nu}-h_{\alpha\beta;\mu\nu}-h_{\mu\nu;\alpha\beta}\right)\label{second-order_Ricci}
\end{align}
is the second-order Ricci tensor, where a subscript semicolon denotes a covariant derivative compatible with $g_{\mu\nu}$. The second-order numerical variable in Eq.~\eqref{EFE2} is the ``residual'' field $h^{2\res}_{\mu\nu}=h^{2}_{\mu\nu}-h^{2\P}_{\mu\nu}$, defined here in terms of the ``puncture'' field $h^{2\P}_{\mu\nu}$. The puncture is a local expansion of the small object's second-order self-field near the worldline $z^\mu$ (given explicitly in covariant form in Ref.~\cite{Pound-Miller:14}), and it goes  to zero at some finite distance from the worldline. We also require analogous first-order fields related by $h^{1\res}_{\mu\nu}=h^{1}_{\mu\nu}-h^{1\P}_{\mu\nu}$.

These field equations are coupled to the equation of motion
\beq\label{EOM-gravity}
\frac{D^2z^\alpha}{d\tau^2} =  -\frac{1}{2}P^{\alpha\mu}
		\left(g_\mu{}^\delta-h^{\res}_\mu{}^\delta\right)\!\left(2h^\res_{\delta\beta;\gamma}-h^\res_{\beta\gamma;\delta}\right)\!u^\beta u^\gamma,
\eeq
where $\tau$ is proper time in $g_{\mu\nu}$, $u^\mu=\frac{dz^\mu}{d\tau}$ is the four-velocity, $P^{\mu\nu}=g^{\mu\nu}+u^\mu u^\nu$ projects orthogonally to the worldline, $h^\res_{\mu\nu}=\e h^{1\res}_{\mu\nu}+\e^2h^{2\res}_{\mu\nu}$ is the total residual field, and $\e=m/L\ll1$, where $L$ is an external length scale (such as the mass $M$ of the large black hole in an EMRI). Equation~\eqref{EOM-gravity} governs the worldline $z^\mu$ on which $T_{\mu\nu}$ has support and on which the punctures $h^{n\P}_{\mu\nu}$ diverge.

\subsection{Second-order scalar-field model}
In place of the above set of equations, I now consider the following nonlinear scalar field model in Minkowski spacetime.

In analogy with Eqs.~\eqref{EFE1}--\eqref{EFE2}, I adopt the field equations
\begin{align}
\Box\varphi_1 &= -4\pi\rho=:S_1(x;z),\label{phi1}\\
\Box\varphi^\res_2 &= t^{\alpha\beta}\nabla_{\alpha}\varphi_1\nabla_\beta \varphi_1 - \Box\varphi^\P_2=: S_2(x;z),\label{phi2}
\end{align}
where $\Box=(-\partial_t^2+\partial_i\partial^i)$ is the flat-space d'Alembertian and $t^{\mu\nu}$ is a nondynamical coupling tensor given by $t^{\mu\nu}={\rm diag}(1,1,1,1)$ in Cartesian coordinates $(t,x^i)$; this coupling is chosen to mimic the large-$r$ behavior of $\delta^2R_{\alpha\beta}$, as described in Sec.~\ref{2nd-order-psi}. The first-order source is the point charge distribution
\beq
\rho(x;z) = \int_\gamma \frac{\delta^4(x-z(\tau))}{\sqrt{-g}}d\tau.
\eeq
(Throughout this paper, I factor out all powers of the charge $q$ and any other length scales, such that all quantities are dimensionless.) Without loss of generality, I place the charge on an equatorial trajectory $z^\mu(t)=(t,r_p(t),\pi/2,\phi_p(t))$, allowing me to write
\beq\label{rho}
\rho(t,r,\theta^A) = \frac{\delta(r-r_p(t))}{r^2U(t)}\sum_{\ell m}Y^*_{\ell m}(\theta^A_p(t))Y_{\ell m}(\theta^A),
\eeq
where $\theta^A_p(t)=(\pi/2,\phi_p(t))$, $U(t):=u^t(t)=1/\sqrt{1-r_p^2(t)\Omega^2(t)}$, and $\Omega(t):=\frac{d\phi_p}{dt}$. The sum runs over all possible values of $\ell\geq0$, $-\ell\leq m\leq \ell$; unless otherwise stated, all sums in this paper do likewise. Note that the solutions to Eqs.~\eqref{phi1}--\eqref{phi2} inherit a functional dependence on $z^\mu$ from their sources, and we may write them as $\varphi_1(x;z)$ and $\varphi^\res_2(x;z)$.

In analogy with Eq.~\eqref{EOM-gravity}, I couple the field equations to the equation of motion
\beq\label{EOM}
\frac{D^2z^\mu}{d\tau^2} = f_{\rm ext}^\mu+\e f_{1\,\rm self}^\mu+\e^2 f_{2\,\rm self}^\mu,
\eeq
where 
\beq
f_{\rm ext}^\mu=-\frac{U^2}{r_p^2}\delta^\mu_r
\eeq
is a (relativistic) Coulomb-type radial force per unit mass, 
\beq\label{fnself}
f_{n\,\rm self}^\mu=P^{\mu\nu}\partial_\nu\varphi_n^\res
\eeq
is the $n$th-order self-force per unit mass, and $\e\ll1$ is a small parameter analogous to $m/L$. I write the worldline's dependence on $\e$ as $z^\mu(t,\e)$. %i.e. scale all distances by L:=qQ/m. Then epsilon= 

Although these equations involve punctures $\varphi_n^{\P}$ and residual fields $\varphi_n^{\res}$, here I am not concerned with precise definitions of those fields. For the purposes of this paper, they may be chosen in any way that guarantees (i) they possess the same symmetries as the orbit (in a manner that will be made clear below), (ii) $\partial_\mu\varphi_n^\res\not\equiv0$ on $z^\mu$, and (iii) $\Box\varphi^\P_n = S_n + \O(\lambda^0)$, where $\lambda$ is a characteristic distance from $z^\mu$. Requirement (iii) ensures that both Eqs.~\eqref{phi2} and \eqref{fnself} are well defined on the particle. 

This toy model has multiple peculiar features: it involves a preferred reference frame, and unlike a standard scalar charge, the particle does not have a time-varying mass. However, the model is not intended to be in any way physical. Its sole purpose is to mimic the key features of the second-order gravitational problem. In particular, the fields $\varphi_n$ exhibit the same behavior as $h^n_{\mu\nu}$ at large distances, and the worldline inspirals in the same manner. To further simplify the model, I specialize to a quasicircular trajectory, with $\frac{dr_p}{dt}\sim\frac{d\Omega}{dt}\sim\e$, making the setup here even more closely analogous to the gravitational one in Ref.~\cite{Pound:14c}. The consistency of such an orbit with Eq.~\eqref{EOM} is ensured by the requirement (i) mentioned above.

\subsection{Scales in the problem and domains of validity}%: near zone, evolution zone, and far zone}
As described in the introduction, the obstacles we face emerge from an intrinsic feature of the physical scenario: the presence of multiple scales. Motivating the various approximations in later sections requires some understanding of these scales. 

We can identify two important ones. First there is the orbital scale $\sim\e^0$, which is the scale of both the orbital period $2\pi/\Omega$ and the orbital radius $r_p$. Next there is the radiation-reaction timescale $\sim 1/\e$, which is the time needed for the orbital radius to shrink by an amount $\sim \e^0$ due to the dissipative effect of $f^\mu_{\rm self}$. %Finally, there is the total inspiral time $ \gg 1/\e$, which is the time needed for the orbit to go from an infinite orbital radius in the infinite past to a  radius $\sim\e^0$. This last scale is not intrinsic to the dynamics of the system, but is extrinsic, set by the boundary conditions of the inspiraling system we wish to describe: we want the system to be in an infinite box, with no radiation entering the box.

Because the fields propagate along null curves, these timescales are entangled with spatial scales. The first-order field $\varphi_1$ propagates from the particle alone, implying that changes on the timescale $\sim\e^n$ along the particle's worldline introduce changes over distances $\sim\e^n$ from that worldline. However, because the second-order source is noncompact, the second-order field $\varphi_2$ propagates from every point in spacetime, and there is no easy correspondence between the time and space scales. Hence, I simply divide the spatial domain into a near zone, $r\sim \e^0$, and a far zone, $r\gg\e^0$.

%Corresponding to these three scales, I divide the entire spacetime into three regions. First choose some point $z^\mu(t_0)$ on the worldline, with $r_p(t_0)\sim\e^0$. Define the \emph{near zone} to be a region of  size $(t,r)\sim \e^0$ around that point; this is the zone in which we can think of the orbit as approximately circular with radius $r_p(t)\sim r_p(t_0)$. Next define the \emph{evolution zone} to be a region of size $(t,r)\sim1/\e$ containing the near zone; this is the zone in which the slow evolution of the orbital radius becomes significant, changing by an amount $\Delta r_p\sim\e^0$. Finally define the \emph{far zone} to be a region of size $(t,r)\gg 1/\e$ containing $z^\mu(t_0)$; this is the zone in which the orbital radius changes by a large amount of size $\Delta r_p\gg 1$.%unlike the evolution zone's inclusion of the near zone, the far zone does not include the near or evolution zone. 

In the following sections, I introduce sequentially more information from each of these timescales and spatial regions. Section~\ref{near-zone} uses the most natural expansion valid on the orbital timescale and shows what goes wrong when it is taken to be valid outside the near zone. Section~\ref{multiscale} uses a multiscale expansion to enlarge the domain of validity to the radiation-reaction timescale. Finally, Sec.~\ref{matching} incorporates information from the far zone. The end result is an approximation for $\e\varphi_1+\e^2\varphi_2$ that is uniformly accurate through order $\e$ over the radiation-reaction time in both the near and far zones, and accurate through $\e^2$ on the orbital timescale in the near zone. More work would be required to obtain order-$\e^2$ accuracy on large scales; however, the approximation here is sufficient to compute first-order-accurate waveforms for complete inspirals and second-order-accurate local-in-time conservative effects, the two principal types of quantities of interest. %Along the way, the domain of validity of varportions of these zones in which the 

Following the nomenclature in, e.g., Refs.~\cite{Pound:14c,Pound:15b}, I refer to the expansion on the orbital timescale as the Gralla-Wald expansion, after the authors of Ref.~\cite{Gralla-Wald:08}. To help distinguish it from the multiscale expansion, I will use a breve, as in $\GW{\varphi}_n$, to refer to $n$th-order terms in it, and a tilde, as in $\tilde\varphi_n$, to those in the multiscale expansion.
%In defining these three zones, I have referred not only to time scales but to space scales. The reason is that the field equations~\eqref{} are hyperbolic, meaning that information from the source propagates outward at a finite speed. Therefore, time scales over which the orbit changes translate directly into spatial scales over which the field changes.

%%%%%%%%%%%%%%%%%%%%%%%%%%%%%%%%%%%%%%%%%%%%%%%%%%%%%%%
\section{Expansion on the orbital timescale}\label{near-zone}
The Gralla-Wald expansion I use is based on one core premise: over a time $\sim\e^0$, the quasicircular orbit deviates from a precisely circular orbit by only a small amount of size $\ll 1$. Based on this, I suppose that we may expand the worldline as
\beq\label{GW-expansion-z}
z^\mu(t,\e) = \GW{z}^\mu_0(t) + \e \GW{z}^\mu_1(t) + \O(\e^2),
\eeq
where the zeroth-order term is a precisely circular orbit of radius $\GW{r}_0$,
\beq
\GW{z}^\mu_0(t) = (t,\GW{r}_0,\pi/2,\GW{\Omega}_0 t),
\eeq
and $\GW{z}_1^\mu$ is the leading deviation from $\GW{z}_0^\mu$. Here we are not particularly interested in the explicit solutions for $\GW{z}^\mu_n$, but we find from the equation of motion~\eqref{EOM} that the zeroth-order frequency is $\GW{\Omega}_0 = \sqrt{\frac{1}{\GW{r}_0^3}}$, and that the deviations grow as 
\beq\label{scaling}
\GW{r}_1(t)= r_{11}t+r_{10}, \qquad \GW{\phi}_1(t)=\phi_{12}t^2+\phi_{11}t+\phi_{10},
\eeq
for some constants $r_{1k}$ and $\phi_{1k}$. Explicit results can be worked out following Appendix~A of Ref.~\cite{Pound:14c}, but the forms~\eqref{scaling} also follow immediately from the facts that $r_p$ and $\Omega$ are slowly evolving, as $\dot r_p\sim \dot\Omega\sim\e$, and that $\phi_p=\int \Omega dt$.

%\frac{D^2z_1^\mu}{d\tau^2} &= f^\mu_{1\,\rm self}(z_0(\tau);z_0)+r_1\partial_{r_0}f^\mu_{\rm ext}(r_0).
%\end{align}

If we substitute the expansion \eqref{GW-expansion-z} into the fields $\varphi_1(x;z)$, $\rho(x;z)$, et cetera, we obtain 
\begin{align}
\varphi_1(x;z) &= \varphi_1(x;\GW{z}_0) + \e\delta \varphi_1(x;\GW{z}_0,\GW{z}_1)+\O(\e^2),\label{GW-expansion-phi}\\
\rho(x;z) &= \rho(x;\GW{z}_0) + \e\delta \rho(x;\GW{z}_0,\GW{z}_1)+\O(\e^2),\label{GW-expansion-rho}
\end{align}
and so forth, where for a functional $f(x;z)$, $\delta f(x;\GW{z}_0,\GW{z}_1):=\frac{d}{d\lambda}f(x;\GW{z}_0+\lambda \GW{z}_1)\big|_{\lambda=0}$. This gives us new variables to work with,
\begin{align}
\GW{\varphi}_1(x;\GW{z}_0) &= \varphi_1(x;\GW{z}_0), \\
\GW{\varphi}_2(x;\GW{z}_0) &= \varphi_2(x;\GW{z}_0)+\delta\varphi_1(x;\GW{z}_0,\GW{z}_1).
\end{align}
From Eqs.~\eqref{phi1} and \eqref{phi2}, these new variables satisfy 
\begin{align}
\Box\GW{\varphi}_1 &= -4\pi\GW{\rho}=:\GW{S}_1,\label{GWeq1}\\
\Box\GW{\varphi}^\res_2 &= t^{\alpha\beta}\nabla_{\alpha}\GW{\varphi}_1\nabla_\beta \GW{\varphi}_1-4\pi\delta\rho - \Box\GW{\varphi}^\P_2,\label{GWeq2}
\end{align}
where $\GW{\rho}=\rho(x;z_0)$. 

An expansion of this type underlies many calculations in the gravitational self-force literature. In particular, it is at the heart of the analysis of second-order conservative effects in Ref.~\cite{Pound:14c}. It is manifestly prone to growing errors on large timescales, due to the growth in Eq.~\eqref{scaling}, but here I will be more interested in the subtler problems it encounters on large spatial scales. To delineate those problems, I deal with the second-order field's two pieces separately. Writing it as $\GW{\varphi}_2=\GW{\psi}+\delta\varphi_1$, where $\GW{\psi}=\varphi_2(x;\GW{z}_0)$, one can obtain $\delta\varphi_1$ as the retarded solution to
\beq
\Box\delta\varphi_1 = -4\pi\delta\rho, \label{dphi}
\eeq
and the residual part of $\GW{\psi}=\GW{\psi}^\res+\GW{\psi}^\P$ as the retarded solution to
\beq\label{psi-eq}
\Box\GW{\psi}^\res = t^{\alpha\beta}\nabla_{\alpha}\GW{\varphi}_1\nabla_\beta \GW{\varphi}_1 - \Box\GW{\psi}^\P =: \GW{S}_2.
\eeq
Note that the puncture is given simply by $\GW{\psi}^\P=\varphi^\P_2(x;\GW{z}_0)$, and the source by $\GW{S}_2 = S_2(x;\GW{z}_0)$. In words, $\delta\varphi_1$ is the part of $\GW{\varphi}_2$ sourced by the deviation of the charge away from $\GW{z}_0^\mu$, and $\GW{\psi}$ is the part sourced by the nonlinear quantity $S_2$. As we shall see, the problem of secular growth occurs in $\delta\varphi_1$, while the problem of infrared divergence occurs in $\GW{\psi}^\res$ (or equivalently in $\GW{\psi}$).

I discuss the above field equations in sequence: Eq.~\eqref{GWeq1} in Sec.~\ref{1st-order}, Eq.~\eqref{dphi} in Sec.~\ref{2nd-order-dphi}, and Eq.~\eqref{psi-eq} in Sec.~\ref{2nd-order-psi}. 

\subsection{First-order solution}\label{1st-order}

When evaluated at $z^\mu=\GW{z}^\mu_0$, the first-order charge distribution~\eqref{rho} takes a simple form with a discrete set of frequencies $m\GW{\Omega}_0$:
\beq
\GW{\rho} = \frac{\delta(r-\GW{r}_0)}{\GW{U}_0r^2}\sum_{\ell m}N_{\ell m}e^{-im\GW{\Omega}_0 t}Y_{\ell m}(\theta^A),
\eeq
where $\GW{U}_0=1/\sqrt{1-1/\GW{r}_0}$, and I have used $\GW{\theta}^A_0(t)=(\pi/2,\GW{\Omega}_0 t)$ to write
\beq
Y^*_{\ell m}(\GW{\theta}^A_0(t)) = N_{\ell m}e^{-im\GW{\Omega}_0 t}
\eeq
with
\beq
N_{\ell m}:=(-1)^m\sqrt{\frac{2\ell+1}{4\pi}\frac{(\ell-m)!}{(\ell+m)!}}P_\ell^m(0).
\eeq

The retarded solution to Eq.~\eqref{GWeq1} possesses the same frequencies as its source. So we write
\beq\label{phi1-lm}
\GW{\varphi}_1(t,r,\theta^A) = \sum_{\ell m} \GW{\varphi}_{1\ell m}(t,r)Y_{\ell m}(\theta^A), 
\eeq
with 
\beq\label{phi1-lmw}
\GW{\varphi}_{1\ell m}(t,r) = \GW{R}_{1\ell m}(r)e^{-im\GW{\Omega}_0t}.
\eeq
Equation~\eqref{GWeq1} then reads
\begin{align}
\partial^2_r\GW{R}_{1\ell m}+\frac{2}{r}&\partial_r\GW{R}_{1\ell m} + \left[m^2\GW{\Omega}_0^2-\frac{\ell(\ell+1)}{r^2}\right]\GW{R}_{1\ell m} \nonumber\\
		&= -\frac{4\pi }{\GW{U}_0}N_{\ell m}\frac{\delta(r-\GW{r}_0)}{r^2}=:\GW{S}_{1\ell m}. \label{GWeq1-lmw}
\end{align}

I solve Eq.~\eqref{GWeq1-lmw} separately for the $m=0$ and $m\neq0$ modes; note that because of Eq.~\eqref{phi1-lmw}, these two cases respectively correspond to the stationary and nonstationary pieces of the field. In both cases, the retarded solution can be written in terms of homogeneous solutions $\GW{R}^\pm_{\ell m}$ as
\beq\label{GWphi1}
\GW{R}_{1\ell m}(r) = \GW{c}^+_{1 \ell m}(r)\GW{R}^+_{\ell m}(r) + \GW{c}^-_{1 \ell m}(r)\GW{R}^-_{\ell m}(r), 
\eeq
with weighting functions
\begin{align}
\GW{c}^+_{1\ell m}(r) &= \int_0^r \frac{\GW{R}^-_{\ell m}(r')\GW{S}_{1\ell m}(r')}{\GW{W}_{\ell m}(r')}dr',\\
\GW{c}^-_{1\ell m}(r) &= \int_r^\infty \frac{\GW{R}^+_{\ell m}(r')\GW{S}_{1\ell m}(r')}{\GW{W}_{\ell m}(r')}dr' .
\end{align}
The homogeneous solution $\GW{R}^+_{\ell m}$ is regular at infinity and contains no incoming waves, $\GW{R}^-_{\ell m}$ is regular at $r=0$, and $\GW{W}_{\ell m}=\GW{R}^-_{\ell m}\partial_r\GW{R}^+_{\ell m}-\GW{R}^+_{\ell m}\partial_r\GW{R}^-_{\ell m}$ is the Wronskian.

\subsubsection{$m=0$ modes}
First apply Eq.~\eqref{GWphi1} for $m=0$. The appropriate homogeneous solutions are
\begin{align}
\GW{R}^+_{\ell 0} = \frac{1}{r^{\ell+1}},\qquad \GW{R}^-_{\ell 0} = r^{\ell},\label{Rl0}
\end{align}
the Wronskian is 
\begin{align}\label{Wl0}
\GW{W}_{\ell0} = -\frac{2\ell+1}{r^2},
\end{align}
and the regular inhomogeneous solution is
\beq
\GW{R}_{1\ell 0} = \frac{4\pi}{\GW{U}_0}\frac{N_{\ell0}}{2\ell+1}\frac{\GW{r}_<^\ell}{\GW{r}_>^{\ell+1}},\label{GWphi1-m=0}
\eeq
where $\GW{r}_<:={\rm min}(r,\GW{r}_0)$ and $\GW{r}_>:={\rm max}(r,\GW{r}_0)$. 

At this stage, there is no indication that anything has gone wrong. We have obtained a global solution to the first-order equation without difficulty, and it does not exhibit any obvious ill behavior. %However, in the next two sections we shall see how our treatment of the near-zone expansion gives rise to problems at second order. 

\subsubsection{$m\neq0$ modes}
Now apply Eq.~\eqref{GWphi1} for $m\neq0$. The appropriate homogeneous solutions are
\begin{align}
\GW{R}^+_{\ell m}(r) = h^{(1)}_{\ell}(m\GW{\Omega}_0 r), \qquad \GW{R}^-_{\ell m}(r) &= j_{\ell}(m\GW{\Omega}_0 r), \label{Rlm}
\end{align}
where $h^{(1)}_{\ell}$ is the spherical Hankel function of the first kind, and $j_\ell$ is the spherical Bessel function of the first kind. The Wronskian can be simplified to
\beq\label{Wlm}
\GW{W}_{\ell m} = -\frac{1}{im\GW{\Omega}_0 r^2}.
\eeq
Putting these together in Eq.~\eqref{GWphi1}, we get the retarded solution
\beq\label{GWphi1-ret}
\GW{R}_{1\ell m} = \frac{4\pi i}{\GW{U}_0}N_{\ell m}  m \GW{\Omega}_0 j_\ell(m\GW{\Omega}_0 \GW{r}_<)h^{(1)}_{\ell}(m\GW{\Omega}_0 \GW{r}_>).
\eeq

We are mainly interested in the asymptotic behavior of these modes at large $r$. Using the  approximation 
\beq\label{Hankel-asymptotic}
h^{(1)}_\ell(z) = (-i)^{\ell+1} \frac{e^{iz}}{z}+\O(1/z^2),
\eeq
we may write
\beq\label{GWphi1-1/r}
\GW{R}_{1\ell m} = \frac{4\pi (-i)^\ell}{\GW{U}_0} N_{\ell m} j_\ell(m\GW{\Omega}_0 \GW{r}_0)\frac{e^{im\GW{\Omega}_0 r}}{r} +\O(1/r^2).
\eeq
%and hence
%\beq
%\GW{\varphi}_{1} = \frac{4\pi}{u^t_0}i\sum_{\ell m}N_{\ell m} j_\ell(m\Omega_0 r_0)\frac{e^{-im\Omega_0 u}}{r}Y_{\ell m} +\O(1/r^2),
%\eeq
From Eq.~\eqref{phi1-lmw} we see that this leads to the time-domain behavior 
\beq\label{GWphi1-1/r-2}
\GW{\varphi}_{1\ell m} \sim \frac{e^{-im\GW{\Omega}_0u}}{r},
\eeq
where $u=t-r$ is the retarded time coordinate. 

Equation~\eqref{GWphi1-1/r-2} is the expected behavior for an outgoing wave, and as with the $m=0$ modes, superficially, nothing has gone wrong in our calculation. However, the solution contains a significant flaw: the waves in Eq.~\eqref{GWphi1-1/r-2} persist even in the infinite past, $u\to-\infty$. On the Penrose diagram of Minkowski space, the waves reach spatial infinity, not only future null infinity. Contrary to this behavior, the exact solution $\varphi_1$ is asymptotically stationary in the infinite past, since the particle's orbit asymptotes to zero frequency there; hence, the exact solution contains no oscillations at spatial infinity. In other words, even in the range of $t$ that $\GW{\varphi_1}$ is expected to be valid, it is a spatially nonuniform approximation. The problems we encounter at second order will emerge from this nonuniformity.%We can also pull out $e^{im\phi}$ from $Y_{\ell m}$ to write the overall oscillatory factor in the helically symmetric form $e^{im(\phi-\Omega_0 u)}$.

%For comparison, the advanced solution is
%\beq\label{GWphi1-adv}
%\GW{R}^{\rm adv}_{1\ell m} = -\frac{4\pi}{u^t_0}N_{\ell m} i m  \Omega_0 j_\ell(m\Omega_0 r_<)h^{(2)}_{\ell}(m\Omega_0 r_>),
%\eeq
%where $h^{(2)}_\ell$ is the spherical Hankel function of the second kind.

\subsection{Second-order solution:  secular growth in $\delta\varphi_1$}\label{2nd-order-dphi}
The nonuniformity of the Gralla-Wald approximation is best illustrated by the growth in its second-order term $\delta\varphi_1$. As shown in Eq.~\eqref{scaling}, the perturbation $\GW{z}^\mu_1$ to the particle's position grows as $\GW{r}_1\sim t$ and $\GW{\phi}_1\sim t^2$.\footnote{Now that the calculation of $\GW{\varphi}_1$ is complete, I can specify what I meant in Sec.~\ref{model} by the requirement that $\varphi_n^\P$ and $\varphi_n^\res$ carry the symmetries of the orbit. Concretely, $\GW{\varphi}_1^\P$ and $\GW{\varphi}_1^\res$ must have expansions exactly analogous to Eqs.~\eqref{phi1-lm}--\eqref{phi1-lmw}, making $\partial_\mu\GW{\varphi}^\res_1$ constant along $\GW{z}^\mu_0$; and  like $\GW{\varphi}_1$, at $\theta=\pi/2$ they must satisfy $\partial_\theta\GW{\varphi}^\res_1=\partial_\theta\GW{\varphi}^\P_1=0$. These conditions make the leading term in the self-force, $\GW{f}^\mu_{1\rm self}= (g^{\mu\nu}+\GW{u}^\mu_0\GW{u}^\nu_0)\partial_\nu\GW{\varphi}_1^\res$, constant and planar, which is enough to ensure Eq.~\eqref{scaling}. Analogous conditions will be imposed in the two-timescale expansion.} Now I examine the growth this leads to in $\delta\varphi_1$. 

We can calculate the growth in two ways. One way is to directly solve Eq.~\eqref{dphi} by integrating the source $\delta\rho$ against the standard retarded Green's function. The second way is to solve the original equation~\eqref{phi1} for the field $\varphi_1(x;z)$, and only afterward substitute the expansion of $z^\mu$. Because knowledge of the exact solution $\varphi_1(x;z)$ will help illuminate the limitations of $\GW{\varphi}_1$ in Sec.~\ref{2nd-order-psi}, I follow the second route.  

%The exact solution to Eq.~\eqref{phi1-eq} can be written in terms of STF quantities as 
%\beq
%\varphi_1 = \sum_{\ell}\frac{(-1)^\ell}{\ell!} \partial^L \bar\rho_L(u),
%\eeq
%:
%\beq
%\bar\rho_L(u) = \int d^3x' \hat x'^L
%\eeq

\subsubsection{Asymptotics of the exact solution}
To characterize the growth in $\delta\varphi_1$, we need only examine the large-$r$ asymptotics. Hence, I seek an asymptotic solution to Eq.~\eqref{phi1}.

Fortunately, such a solution is ready at hand in the PM literature, most clearly in Ref.~\cite{Damour-Iyer:91}. In Cartesian coordinates $(t,\vec{x})=(t,x^a)$, the retarded solution to Eq.~\eqref{phi1}, for any compact source distribution $\rho(t,\vec x)$, can be written as
\beq\label{phi1-PW}
\varphi_1 = \sum_{\ell}\frac{(-1)^\ell}{\ell!}\partial_L\frac{\bar\rho^L(t-r)}{r}.
\eeq
Here $L=i_1\cdots i_\ell$ is a multi-index, $\partial_L=\partial_{i_1}\cdots\partial_{i_\ell}$, and summation over the $\ell$ contracted indices is implied; this notation will recur in later sections. $\bar\rho^L(u)$ is the $\ell$th moment of a certain weighted time-average of $\rho(t,\vec x)$,
\beq
\bar \rho^L(u) = \int \bar \rho_\ell(u,\vec x') \hat x'^L d^3x',
\eeq
where $\hat x^L = x^{\langle i_1}\cdots x^{i_\ell\rangle}$ is a symmetric-trace-free (STF) product,
\beq
\bar \rho_\ell(u,\vec x') = \int_{-1}^1 \delta_\ell(z)\rho(u+r'z,\vec x')dz,
\eeq 
and $\delta_\ell(z) = \frac{(2\ell+1)!!}{2^{\ell+1}\ell!}(1-z^2)^\ell$.

It will suffice to pick off the leading term in Eq.~\eqref{phi1-PW}. To do so, I introduce the outward-pointing unit vector $n^i=\frac{x^i}{r}$. In terms of this unit vector, a spatial derivative acts on $1/r$ as $\partial_i \frac{1}{r}=-\frac{n_i}{r^2}$, increasing the power of $1/r$, while a spatial derivative acts on $\bar\rho_L(u)$ as $\partial_i\bar\rho_L(u) = -n_i \frac{d\bar\rho_L}{du}$, having no effect on the overall power of $r$. Similarly, a spatial derivative acting on $n_i$ adds a power of $1/r$, and so we can write the large-$r$ behavior of Eq.~\eqref{phi1-PW} as
\beq\label{phi1-1/r}
\varphi_1 = \frac{1}{r}\sum_{\ell}\frac{\nhat_L}{\ell!} \frac{d^\ell}{du^\ell}\bar\rho^L(u) + \O(1/r^2),
\eeq
where $\nhat^L = n^{\langle i_1}\cdots n^{i_\ell\rangle}$. This can be expressed in terms of ordinary spherical harmonics by moving the factor of $\nhat_L$ inside the integral over $\vec x'$ and utilizing the identity~\cite{Blanchet-Damour:86}
\beq
n_L n'^L = \nhat_L\nhat'^L =\sum_m \frac{4\pi \ell!}{(2\ell+1)!!}Y^*_{\ell m}(\theta'^A)Y_{\ell m}(\theta^A).
\eeq
Equation~\eqref{phi1-1/r} then becomes
\beq\label{phi1-1/rlm}
\varphi_1 = \frac{4\pi}{r}\sum_{\ell m}\frac{ Y_{\ell m}}{(2\ell+1)!!}\frac{d^\ell}{du^\ell}\!\int\! \bar\rho_{\ell m}(u,r')r'^{\ell+2}dr' + \O(1/r^2),
\eeq
where $\bar\rho_{\ell m}(u,r') = \int d\Omega' Y^*_{\ell m}(\theta'^A)\bar\rho_\ell(u,\vec x')$.% with $\vec{x} = r \vec{n}(\theta^A)$.

Equation~\eqref{phi1-1/rlm} is valid for any compact source distribution $\rho$. For our point source~\eqref{rho}, we can eliminate the integrals over $\theta'^A$ and $r'$ to obtain 
\begin{align}\label{phi1-1/r-v2}
\varphi_1 &= \frac{4\pi}{r}\sum_{\ell m k}\frac{N_{\ell m} Y_{\ell m}}{(2\ell+1)!!}\frac{d^\ell}{du^\ell}\!\int_{-1}^1\! dz \frac{ \delta_\ell(z)r_{(k)}^\ell e^{-im\phi_p(u_{(k)})}}{U(u_{(k)})|1-z\dot r_p(u_{(k)})|} \nonumber\\
&\quad + \O(1/r^2),
\end{align}
where $r_{(k)}=r_{(k)}(u,z)$ is the $k$th solution to $r_{(k)}=r_p(u+r_{(k)}z)$, and $u_{(k)}=u+z r_{(k)}$.
%Substituting this into Eq.~\eqref{phi1-1/rlm} gives us
%\begin{align}\label{phi1-1/r-v2}
%\varphi_1 &= \frac{4\pi}{r}\sum_{\ell m}\frac{N_{\ell m} Y_{\ell m}}{(2\ell+1)!!} \frac{d^\ell}{du^\ell}\!\!\left[\frac{r_p^{\ell}(u)}{u^t(u)}e^{-im\phi_p(u)}\right]  + \O(1/r^2).
%\end{align}

Obviously this solution is not complete in itself. It depends in a complicated way on the worldline variables $r_p$, $\phi_p$, and $U$. And through the equation of motion~\eqref{EOM}, those variables reciprocally depend on $\varphi_1$. However, Eq.~\eqref{phi1-1/r-v2} illuminates the basic structure of the exact solution, and it provides enough information for my purposes in the sections below.

% To compare with the results of the previous section, I express the multipole expansion in terms of spherical harmonics $Y_{\ell m}$,  
%\beq
%\varphi_1 = \sum_{\ell}\frac{(-1)^\ell}{\ell!}\sum_{k=1}^\ell \partial^k_r \frac{\rho_\ell m(u)}{r}.
%\eeq

\subsubsection{Growing errors in the Gralla-Wald expansion}
We may now find $\delta\varphi_1$ by substituting the expansion of $z^\mu$ into the exact solution~\eqref{phi1-1/r-v2}.% Before doing so, we may immediately note the solution~\eqref{phi1-1/r-v2} depends on the worldline evaluated at the retarded time $u=t-r$, not at time $t$. This implies that when we expand the worldline as in Eq.~\eqref{GW-expansion-z}, the growth in $z^\mu_1$ at large $t$ along the worldline generates growth in $\delta\GW{\varphi}_1$ at large $r$ away from the worldline. 

Using Eq.~\eqref{GW-expansion-z} [with Eq.~\eqref{scaling}], we find that $r_{(k)}=r_p(u_{(k)})\approx r_p(u+\GW{r}_0z)$ has only one solution: 
\beq
r_{(k)} =\GW{r}_0+\e[r_{11}(u+\GW{r}_0z)+r_{10}]+\O(\e^2). %We also have that $\dot r_p(u_{(k)}) = \e r_{11}+\O(\e^2)$ is independent of $u_{(k)}$ at leading order. 
\eeq
Equation~\eqref{phi1-1/r-v2} then becomes
%\begin{align}
%\frac{r_p^{\ell}(u)}{u^t(u)}e^{-im\phi_p(u)} &= \frac{\GW{r}_0^{\ell}}{\GW{u}_0^t}e^{-im\GW{\Omega}_0 u}[1+ \e (a_{\ell m}u+b_{\ell m}u^2)]\nonumber\\
%				&\quad +\O(\e^2u^3)
%\end{align}
%for some constants $a_{\ell m}$ and $b_{\ell m}$. The $u^2$ term comes from the expansion of $\phi_p(u)$, meaning that for $m=0$ Eq.~\eqref{} becomes
%\beq
%\frac{r_p^{\ell}(u)}{u^t(u)} = \frac{\GW{r}_0^{\ell}}{\GW{u}_0^t}[1+ \e a_{\ell 0}u+\O(\e^2u^2)].
%\eeq
%Combining these results in Eq.~\eqref{} returns the expansion
\begin{align}\label{phi1-expansion}
\varphi_1 &= \frac{4\pi}{r}\frac{1}{\GW{U}_0}\bigg\{N_{00}Y_{00}[1+\e(a_{00}+b_{00}u)]+\tfrac{1}{3}\e N_{10} Y_{10}b_{10}\nonumber\\
&\quad +\!\sum_{\ell, m\neq0}\!\!\frac{N_{\ell m}Y_{\ell m}}{(2\ell+1)!!}(-i m\GW{\Omega}_0\GW{r}_0)^\ell e^{-i m\GW{\Omega}_0 u}\nonumber\\
&\quad \times\!\!\int_{-1}^1\!dz\, \delta_\ell(z) e^{-im\GW{\Omega}_0\GW{r}_0z}[1+\e (a_{\ell m}+b_{\ell m}u+c_mu^2)]\!\bigg\}\nonumber\\
&\quad +\O(1/r^2,\e^2),
\end{align}
where for $m\neq0$, $a_{\ell m}$, $b_{\ell m}$, and $c_m$ are quadratic, linear, and zeroth-order polynomials in $z$, respectively, with coefficients that depend on $\GW{r}_0$; for $m=0$, the integrals over $z$ have been evaluated and absorbed into $a_{\ell0}$ and $b_{\ell0}$. The first line in Eq.~\eqref{phi1-expansion} is the only contribution of the $m=0$ modes, and the $u^2$ term in the $m\neq0$ modes comes entirely from $\GW{\phi}_1(u)$.

Equation~\eqref{phi1-expansion} is a more explicit form of the expansion~\eqref{GW-expansion-phi}. Hence, we expect the order-$\e^0$ term to precisely recover the zeroth-order term $\GW{\varphi}_1$ in Eqs.~\eqref{GWphi1-1/r} and \eqref{GWphi1-m=0}. We can verify that this is the case by using the identity~\cite{Damour-Iyer:91}
\beq\label{delta_ell}
\delta_\ell(z) = \frac{(2\ell+1)!!}{2\pi}\int_{-\infty}^\infty\! dy\, e^{iyz}\frac{j_\ell(y)}{y^\ell},
\eeq
evaluating the integral over $z$, and then using the identity 
\beq\label{j-integral}
\frac{1}{\pi}\int_{-\infty}^\infty\! dy\, j_0(y-y_0)\frac{j_\ell(y)}{y^\ell} = \frac{j_\ell(y_0)}{(y_0)^\ell}.
\eeq

The order-$\e$ term in Eq.~\eqref{phi1-expansion} is our sought-after  field $\delta\varphi_1$. We may write it as
\begin{align}
\delta\varphi_1 &= \frac{1}{r}\Bigg[ \sum_{\ell, m\neq0}( a'_{\ell m}+ b'_{\ell m}u+ c'_{\ell m}u^2)e^{-i m\GW{\Omega}_0u}Y_{\ell m} \nonumber\\
&\quad  +( a'_{00}+ b'_{00}u)Y_{00} +  b'_{10}Y_{10}\Bigg]+\O(1/r^2),
\end{align}
where constant factors and the integration over $z$ have been absorbed into the constants $ a'_{\ell m}$, $ b'_{\ell m}$, and $ c'_{\ell m}$.

At fixed $u$, near future null infinity $\delta\varphi_1$ falls off in the correct way, as $1/r$; in fact, $\GW{\varphi}_1$ uniformly approximates $\varphi_1$ at fixed $u$. However,  $\delta\varphi_1$ exhibits two types of problems on large scales.  At large retarded times $u$, both in the past and future, it grows large. Moreover, the expansion~\eqref{GW-expansion-phi} fails completely near spatial infinity: at fixed $t$, $\delta\varphi_1$ grows linearly with $r$. 

In Sec.~\ref{multiscale}, I show that these pathologies are eliminated with the use of a multiscale expansion, in which the outgoing waves fall off appropriately in the infinite past and, unlike in $\GW{\varphi}_1$, no waves reach spatial infinity.

\subsection{Second-order solution: infrared divergence in $\GW{\psi}$}\label{2nd-order-psi}

I now turn to the second major issue at second order: the divergent integrals that appear in the retarded solution to Eq.~\eqref{psi-eq}. Like the spurious secular growth, this will be linked to the ill behavior of $\GW{\varphi}_1$ at spatial infinity.

\subsubsection{Harmonic decomposition}\label{S2 harmonic decomposition}
From all appearances, we can solve Eq.~\eqref{psi-eq} using exactly the same method as we did the first-order equation~\eqref{GWeq1}, decomposing the field  $\GW{\psi}$ and source $\GW{S}_2$ into $\ell m \omega$ modes and then solving mode by mode. 

Indeed, we can readily write
\beq\label{S2-decomposition}
\GW{S}_2 = \sum_{\ell m}\GW{S}_{2\ell m}(r;\GW{r}_0)e^{-im\GW{\Omega}_0t}Y_{\ell m}
\eeq
by substituting the expansion~\eqref{phi1-lm}--\eqref{phi1-lmw} into the source $t^{\mu\nu}\nabla_\mu\GW{\varphi}_1\nabla_\nu\GW{\varphi}_1$. Doing so leads to an expansion of the form 
\begin{align}
t^{\mu\nu}\nabla_\mu\GW{\varphi}_1\nabla_\nu\GW{\varphi}_1 &=\!\sum\!\mathcal{D}^{\ell m}_{\ell'm'\ell''m''}\GW{R}_{1\ell'm'}\GW{R}_{1\ell''m''}e^{-im\GW{\Omega}_0t}Y_{\ell m},
\end{align}
where the sum runs over all $\{\ell,\,m,\,\ell',\,m',\,\ell'',\,m''\}$, and $\mathcal{D}^{\ell m}_{\ell'm'\ell''m''}$ is a bilinear differential operator acting on $\GW{R}_{1\ell'm'}$ and $\GW{R}_{1\ell''m''}$; the explicit expression for this expansion, derived in Appendix~\ref{coupling formula}, is given by Eq.~\eqref{S2lm}.  We can also decompose the puncture part of the source, $-\Box\GW{\psi}^\P$, into $\ell m\omega$ modes by first decomposing the puncture itself as $\GW{\psi}^\P = \sum_{\ell m}\GW{R}^\P_{\ell m}(r;\GW{r}_0)e^{-im\GW{\Omega}_0 t}Y_{\ell m}$; the part of $\GW{\varphi}^\P_2$ that does not admit such a decomposition is entirely contained in $\delta\varphi_1$. However, we are interested in the large-$r$ behavior of $\GW{S}_{2\ell m}$, and in that region $\GW{\psi}^\P$ vanishes, allowing us to write
\beq\label{S2lm-compact}
\GW{S}_{2\ell m} = \sum_{\ell'm'}\sum_{\ell''m''} \mathcal{D}^{\ell m}_{\ell'm'\ell''m''}\GW{R}_{1\ell'm'}\GW{R}_{1\ell''m''}.
\eeq

Just as at first order, the retarded solution possesses the same set of frequencies as the source:
\beq\label{psi2lm}
\GW{\psi}^\res = \sum_{\ell m} \GW{\psi}_{\ell m}(t,r)Y_{\ell m}, 
\eeq
with 
\beq\label{phi2-lmw}
\GW{\psi}_{\ell m}(t,r) = \GW{R}_{2\ell m}(r)e^{-im\GW{\Omega}_0t}.
\eeq
Because we will be interested in the large-$r$ behavior, this will serve as a mode of both $\GW{\psi}$ and $\GW{\psi}^\res$. Following precisely the same procedure as in Sec.~\ref{1st-order}, I write the coefficients $\GW{R}_{2\ell m}$ as 
\beq\label{GWphi2}
\GW{R}_{2\ell m}(r) = \GW{c}^+_{2 \ell m}(r)\GW{R}^+_{\ell m}(r) + \GW{c}^-_{2 \ell m}(r)\GW{R}^-_{\ell m}(r), 
\eeq
with weighting functions
\begin{align}
\GW{c}^+_{2\ell m}(r) &= \int_0^r \frac{\GW{R}^-_{\ell m}(r')\GW{S}_{2\ell m}(r';\GW{r}_0)}{\GW{W}_{\ell m}(r')}dr',\\
\GW{c}^-_{2\ell m}(r) &= \int_r^\infty \frac{\GW{R}^+_{\ell m}(r')\GW{S}_{2\ell m}(r';\GW{r}_0)}{\GW{W}_{\ell m}(r')}dr',
\end{align}
and with precisely the same homogeneous solutions $\GW{R}^\pm_{\ell m}$ as in Sec.~\ref{1st-order}.

If the source~\eqref{S2lm-compact} is correct everywhere in spacetime, then Eq.~\eqref{GWphi2} [with Eqs.~\eqref{psi2lm}--\eqref{phi2-lmw}] is the \emph{unique} retarded solution to Eq.~\eqref{psi-eq}; it may alternatively be derived directly from the standard retarded Green's function for the wave operator $\Box$. We shall now see that this unique solution is ill defined, implying that the source~\eqref{S2lm-compact} is \emph{not} correct everywhere in spacetime.

\subsubsection{Asymptotic behavior}
The asymptotic behavior of the source is readily inferred from its explicit expression~\eqref{S2lm} and the asymptotic behaviors in Eqs.~\eqref{GWphi1-1/r} and \eqref{GWphi1-m=0}. Noting that the coupling of modes in Eq.~\eqref{S2lm} imposes $m'+m''=m$, we see that the most slowly decaying terms in $\GW{S}_{2\ell m}$ behave as 
\beq
\frac{e^{im'\GW{\Omega}_0 r}}{r}\frac{e^{im''\GW{\Omega}_0 r}}{r}=\frac{e^{i(m'+m'')\GW{\Omega}_0r}}{r^2}=\frac{e^{im\GW{\Omega}_0r}}{r^2}; 
\eeq
these contributions originate both from $t$-derivative terms of the form $\partial_t \frac{e^{-im'\GW{\Omega}_0 (t-r)}}{r}\partial_t \frac{e^{-im''\GW{\Omega}_0 (t-r)}}{r}$ and from $r$-derivative terms of the form $\frac{\partial_r e^{-im'\GW{\Omega}_0 (t-r)}}{r}\frac{\partial_r e^{-im''\GW{\Omega}_0 (t-r)}}{r}$. Only the oscillatory, $m\neq0$ modes in $\GW{\varphi}_1$ contribute to the $1/r^2$ piece of the source; the stationary modes~\eqref{GWphi1-m=0} contribute only at higher orders in $1/r$. 

This asymptotic behavior guided my choice of toy model source, as it replicates that of the gravitational source $\delta^2R_{\mu\nu}$.  Unlike $\delta^2R_{\mu\nu}$, the most natural source $(\varphi_1)^2$ would have included $1/r^2$ terms coming from stationary modes of $\varphi_1$. A source proportional to $\partial^\mu\varphi_1\partial_\mu\varphi_1$ would have avoided those terms, but it would also have contained no $1/r^2$ terms at all, since the potential $1/r^2$ terms would be $\propto\frac{\partial^\mu e^{-im'u}}{r}\frac{\partial_\mu e^{-im''u}}{r}\propto\frac{\partial^\mu u\partial_\mu u}{r^2}=0$.\footnote{Sources such as $(\partial_t\varphi_1)^2$ were rejected primarily because they would not allow me to present the method of derivation in Appendix~\ref{coupling formula}, which is used in the decomposition of $\delta^2R_{\mu\nu}$ to be presented in a followup paper.}

Now, one can straightforwardly check that all subleading terms in the source, falling off faster than $1/r^2$, generate well-behaved outgoing waves of the form $\sim e^{-im\GW{\Omega}_0 u}/r$ near future null infinity. Hence, only the leading falloff is of interest here. So let us write the source modes as
\beq\label{S2lm-1/r2}
\GW{S}_{2\ell m} = \frac{\GW{S}^{(-2)}_{2\ell m}e^{im\GW{\Omega}_0r}}{r^2}+\O(1/r^3),
\eeq
where $\GW{S}^{(-2)}_{2\ell m}$ is a ($\GW{r}_0$-dependent) constant. Let $\GW{\psi}^\P=0$ beyond $r=r^+$, and let $\GW{R}^{(-2)}_{2\ell m}$ be the part of the solution sourced by $\frac{\GW{S}^{(-2)}_{2\ell m}e^{im\GW{\Omega}_0r}}{r^2}$ at points $r>r^+$. Explicitly,
\begin{align}
\GW{R}^{(-2)}_{2\ell m}(r) &= \GW{S}^{(-2)}_{2\ell m}\left[\GW{R}^+_{\ell m}(r)\int_{r^+}^r \frac{\GW{R}^-_{\ell m}(r')e^{im\GW{\Omega}_0r'}}{r'^2\GW{W}_{\ell m}(r')}dr'
														 \right.\nonumber\\
									&\quad + \left.\GW{R}^-_{\ell m}(r)\int_r^\infty \frac{\GW{R}^+_{\ell m}(r')e^{im\GW{\Omega}_0r'}}{r'^2\GW{W}_{\ell m}(r')}dr'\right]. \label{R-22lm}
\end{align}
At large $r$, the total solution then has the form 
\beq
\GW{R}_{2\ell m}=\GW{R}^{(-2)}_{2\ell m}+\frac{C_{\ell m}e^{im\GW{\Omega}_0r}}{r}+\O(1/r^2)
\eeq
for some $r^+$-dependent constant $C_{\ell m}$. %We are primarily interested in the integral that extends to infinity, the only part of the solution that draws information from large distances.

\subsubsection{$m\neq0$ modes}
As we shall see in the next section, the $m=0$ modes in Eq.~\eqref{R-22lm} are the most problematic. However, apparent irregularities also arise in the oscillatory, $m\neq0$ modes.

For $m\neq0$, Eq.~\eqref{R-22lm} reads
\begin{align}
\GW{R}^{(-2)}_{2\ell m} &= -im\GW{\Omega}_0 \GW{S}^{(-2)}_{2\ell m}\left[h^{(1)}_{\ell}(\bar r)\int_{r^+}^r\! j_\ell(\bar r')e^{i\tilde r'}dr'\right. \nonumber\\
							&\quad \left.+j_\ell(\bar r)\int_r^\infty\! h^{(1)}_{\ell}(\bar r')e^{i\bar r'}dr'\right],
\end{align}
where $\bar r'=m\GW{\Omega}_0 r'$. At large $r$, $h^{(1)}_{\ell}$ behaves as shown in Eq.~\eqref{Hankel-asymptotic}, and $j_\ell$ as
\beq
j_\ell(z)=\begin{cases}(-1)^{\ell/2}\displaystyle\frac{\sin(z)}{z}+\O(1/z^2) & \text{for even $\ell$}\\
 (-1)^{(\ell+1)/2}\displaystyle\frac{\cos(z)}{z}+\O(1/z^2) & \text{for odd $\ell$}.
\end{cases}
\eeq
These asymptotic expressions show that 
\begin{align}
\int_{r^+}^r\! j_\ell(\tilde r')e^{i\tilde r'}dr' &= \frac{i^{\ell+1}\ln r}{2m\GW{\Omega}_0}+\O(r^0),\\
\int_r^\infty\! h^{(1)}_{\ell}(\tilde r')e^{i\tilde r'}dr' &= \frac{(-1)^\ell e^{2im\GW{\Omega}_0r}}{2m^2\GW{\Omega}_0^2r}+\O(1/r^2).
\end{align}
Hence,
\beq\label{R-22lmneq0}
\GW{R}^{(-2)}_{2\ell m} = \frac{(C'_{\ell m}+\GW{S}^{(-2)}_{2\ell m}\ln r)e^{im\GW{\Omega}_0 r}}{2im\GW{\Omega}_0r}+\O(r^{-2}\ln r),
\eeq
for some constant $C'_{\ell m}$.

While this behavior is not obviously problematic, the solution does exhibit some irregularity in the $\frac{\ln r}{r}$ term. In the gravitational problem, such terms violate asymptotic flatness at null infinity. However, the matching procedure in Sec.~\ref{matching} will show that they are actually the correct behavior in the large-$r$ region of the near zone. Furthermore, such terms are well known in PM theory~\cite{Blanchet-Damour:86,Blanchet:87}, where they are a consequence of the metric perturbation deforming the light cones along which solutions to the wave equation propagate. In that context, they are removable via a transformation to an asymptotically regular gauge~\cite{Blanchet:87}.

\subsubsection{$m=0$ modes}
I now specialize to the more problematic, $m=0$ case; this is where the infrared divergence occurs. Since $m'+m''=m=0$, terms of the form $\frac{e^{-i(m'+m'')\GW{\Omega}_0r}}{r^2}$ become simply $1/r^2$, and Eq.~\eqref{S2lm-1/r2} becomes
\beq\label{S2l0}
\GW{S}_{2\ell0} = \frac{\GW{S}^{(-2)}_{2\ell 0}}{r^2}+\O(1/r^3).
\eeq
This stationary-in-time and nonoscillatory-in-$r$ piece of the $1/r^2$ source is generated entirely by the beating of waves of opposite phase, $\frac{e^{im'\GW{\Omega}_0 u}}{r}$ and  $\frac{e^{-im'\GW{\Omega}_0 u}}{r}$. It is the cause of our problems.

Substituting Eqs.~\eqref{Rl0} and \eqref{Wl0} into Eq.~\eqref{R-22lm} yields 
\begin{align}
\GW{R}^{(-2)}_{2\ell 0} &= -\frac{1}{r^{\ell+1}}\GW{S}^{(-2)}_{2\ell0}\int_{r^+}^r \frac{r'^\ell}{2\ell+1}dr' \nonumber\\
									&\quad -r^\ell \GW{S}^{(-2)}_{2\ell0}\int_r^\infty \frac{r'^{-\ell-1}}{2\ell+1}dr'. \label{R-22l0}
\end{align}

First consider the $\ell>0$ modes. Equation~\eqref{R-22l0} evaluates to
\beq\label{R-22lneq00}
\GW{R}^{(-2)}_{2\ell 0}  = -\frac{\GW{S}^{(-2)}_{2\ell0}}{\ell(\ell+1)}+\O(1/r^{\ell+1}).
\eeq
In words, every $m=0$, $\ell>0$ mode approaches a constant at large $r$. In the gravity problem, this corresponds to a lack of asymptotic flatness, signaling either a poorly behaved gauge or a physically ill-behaved approximation. In either case, the behavior appears problematic. 

But the essential problem arises in the $\ell=0$ mode. For this mode, Eq.~\eqref{R-22l0} evaluates to
\begin{align}
%\GW{R}^{(-2)}_{2\ell 0}  &= \frac{C_{00}-r^+S^{(-2)}_{200}}{r} + (1-\ln r)S^{(-2)}_{200}\nonumber\\
%									&\quad +S^{(-2)}_{200}\lim_{r'\to\infty}\ln r'.\label{R-2200}
\GW{R}^{(-2)}_{2\ell 0}  &= \left(\frac{r^+}{r}-1+\lim_{{\cal R}\to\infty}\ln\frac{r}{{\cal R}}\right)\GW{S}^{(-2)}_{200}.\label{R-2200}
\end{align}
The final term is infinite; the solution diverges at all values of $r$. Said another way, the retarded solution simply does not exist. 

In Sec.~\ref{matching}, I cure this divergence with the matching procedure. The same procedure also shows, as with the $m\neq0$ modes, that the suspicious behavior in Eq.~\eqref{R-22lneq00} is actually correct.

%\begin{align}
%S^{-2}_{2\ell 0} &=  \sum_{\ell'm'>0}\sum_{\ell''}\mathcal{D}^{\ell 0}_{\ell'm'\ell''-m'}\GW{\varphi}_{1\ell'm'}\GW{\varphi}_{1\ell''-m'}
%\end{align}

%\begin{align}
%S_{200} &=  \sum_{\ell'}\mathcal{D}^{\ell 0}_{\ell'm'\ell'-m'}\GW{\varphi}_{1\ell'm'}\GW{\varphi}_{1\ell'-m'}
%\end{align}

%In expressing the solution as~\eqref{}, we have assumed a particular asymptotic behavior for the second-order field, and we have assumed that we can safely use the source $S$ constructed from the first-order field even in the far zone. As we shall see in the following sections, both of these assumptions are incorrect.

\subsubsection{Comparison with behavior of exact source $S_{2\ell m}(x;z)$}
%Based on the results for $\delta\varphi_1$ in Sec.~\ref{}, one might hypothesize that working in $(u,r)$ coordinates would eliminate the infrared divergence in $\psi$. Unfortunately, the growing errors in $\delta\varphi_1$ are not directly related to the divergent integral. 

We can discern the origin of the infrared divergence by comparing $S_{2}(x;z_0)$ to the exact source $S_{2}(x;z)$.

To construct $S_{2}(x;z)$, return to the expression for $\varphi_1$ in Eq.~\eqref{phi1-1/r-v2}. If $|u|\lesssim1/\e$, then $r_{(k)}\sim r_p\sim \Omega\sim U\sim \e^0$ inside the integral. For our quasicircular orbit, we also have $\dot\Omega\sim\dot r_p\sim\dot U\sim\e$ and $\dot\phi_p=\Omega\sim\e^0$. Combining these scalings allows us to write $r_{(k)}=r_p(u_{(k)})=r_p(u)+\O(\e)$, $U(u_{(k)})=U(u)+\O(\e)$, $\Omega(u_{(k)})=\Omega(u)+\O(\e)$, and 
\beq
\phi_p(u_{(k)})=\phi_p(u)+\Omega(u)r_p(u)z+\O(\e).
\eeq
We may also evaluate the $u$ derivatives in Eq.~\eqref{phi1-1/r-v2} using the fact that each additional derivative of $r_p(u)$, $\Omega(u)$, and $U(u)$ introduces an additional factor of $\e$, and we can perform the $z$ integration using Eqs.~\eqref{delta_ell} and \eqref{j-integral}. For the $m\neq0$ modes, the end result is
\begin{align}
\varphi_{1\ell m} &= \frac{4\pi (-i)^\ell}{U(u)} N_{\ell m} j_\ell[m\Omega(u) r_p(u)]\frac{e^{-im\phi_p(u)}}{r} \nonumber\\
						&\quad +\O(\e,1/r^2).\label{phi1-uniform}
\end{align}
Unlike $\GW{\varphi}_1$, this is a uniform approximation. It has the simple form of an oscillation (on the orbital timescale) with an amplitude that varies slowly (on the radiation-reaction timescale). If re-expanded on the orbital timescale, it recovers the result for $\GW{\varphi}_1$ in Eq.~\eqref{GWphi1-1/r-2}. But importantly, the amplitude of this wave, unlike that in Eq.~\eqref{GWphi1-1/r-2}, decays toward zero at large past times. We can see this from the facts that $\Omega(u) r_p(u)=r_p(u)^{-1/2}+\O(\e)$ for a quasicircular orbit, that $r_p(u\to-\infty)=\infty$, and that $j_\ell(x\to0)=0$. This is the correct behavior in our physical scenario, in which the particle began infinitely far away, radiating infinitely weak radiation.
  
If we construct $S_{2}(x;z)$ from the field~\eqref{phi1-uniform}, we arrive at 
\beq\label{S2-uniform}
S_{2} = \sum_{\ell m}S_{2\ell m}(r;r_p(u),\Omega(u))e^{-im\phi_p(u)}Y_{\ell m}+\O(\e),
\eeq
where the $\O(\e)$ terms are uniformly small on the radiation-reaction timescale. In particular, we have
\beq
S_{2\ell0} = \frac{S^{(-2)}_{2\ell 0}(u)}{r^2}+\O(\e,1/r^3),
\eeq
where $S^{(-2)}_{2\ell 0}(u)$ is identical to the constant $\GW{S}^{(-2)}_{2\ell 0}$ appearing in Eq.~\eqref{S2l0}---with the replacements $\GW{r}_0\to r_p(u)$ and $\GW{\Omega}_0\to\Omega(u)$. Because  $S^{(-2)}_{2\ell 0}(u)$ is proportional to the square of the amplitude in Eq.~\eqref{phi1-uniform}, the source decays to 0 as $u\to-\infty$, as did the amplitude of $\varphi_{1\ell m}$. Moreover, at fixed $t$, we have $S^{(-2)}_{2\ell 0}(t- r)\to 0$ as $r\to\infty$. Hence, if we had used a uniform approximation to $\varphi_1$, we would have a slowly varying source in the integral~\eqref{R-22l0}, and the integral would have converged. 

The cause of the infrared divergence is now apparent: through the noncompact source $\GW{S}_2$, the second-order retarded solution~\eqref{GWphi2} draws information from $\GW{\varphi}_1$ all the way to spatial infinity, and so it is sensitive to $\GW{\varphi}_1$'s inaccuracy there. Given their nonuniformity, the equations of the Gralla-Wald expansion should instead be solved only in the near zone, on an $\e$-independent domain $r\in[0,{\cal R}]$. I do precisely this in the multiscale expansion in the next section. However, we then need a way to correctly specify boundary conditions at $r={\cal R}$. The matching procedure in Sec.~\ref{matching} provides those boundary conditions.
%Physically, the source~\eqref{} is incorrect because it is identical to the one that would be generated if the particle were on a circular orbit eternally. At larger and larger values of $r$, it samples larger and larger past retarded times, and it should be decaying at those large values of $r$ because the radiation producing it should be decaying at those large past times. 

%%%%%%%%%%%%%%%%%%%%%%%%%%%%%%%%%%%%%%%%%%%%%%%%%%%%%%%
\section{Multiscale expansion}\label{multiscale}
% In all three cases, I follow what is standard practice in black hole perturbation theory: I treat the field equations~\eqref{GWeq1}, \eqref{dphi},\eqref{psi-eq} as globally valid, not merely valid in the near zone, and I seek their global retarded solutions. This would be the correct thing to do if the expansions~\eqref{GW-expansion-z}, \eqref{GW-expansion-phi}, and \eqref{GW-expansion-rho} were uniform; however, the expansions are \emph{not}, and this shall give rise to our problems.\footnote{By uniformity I mean that order symbols such as $\O(\e^n)$ remain valid even if the evaluation point $x^\mu$ depends on $\e$.}

We have now diagnosed the problems but have yet to cure them. In this section I begin that process by presenting a systematic multiscale method of obtaining the complete solution on the radiation-reaction time in the near zone. Section~\ref{multiscale-motion} describes the expansion of the equation of motion, Sec.~\ref{multiscale-field} describes the expansion of the field equations, and Sec.~\ref{multiscale-summary} summarizes the practical combination of the two. The solution is determined up to boundary conditions at large $r$, which are found in Sec.~\ref{matching}.

\subsection{Expansion of the worldline}\label{multiscale-motion}
A multiscale expansion~\cite{Kevorkian-Cole:96} assumes that a function $f(\lambda,\e)$ of variables $\lambda$ and $\e\ll1$ can be uniformly approximated by a series of the form $\sum_n \e^n \tilde f_n(\lambda_{\rm fast}, \lambda_{\rm slow})$, where $\lambda_{\rm fast}(\lambda,\e)\sim\e^0\lambda$ is a ``fast time'', $\lambda_{\rm slow}(\lambda,\e)\sim\e\lambda$ is a ``slow time'', and $\tilde f_n(\lambda_{\rm fast}, \lambda_{\rm slow})\sim\e^0$. When solving a differential equation for $f(\lambda,\e)$, one substitutes $f(\lambda,\e)=\sum_n \e^n \tilde f_n(\lambda_{\rm fast}, \lambda_{\rm slow})$, applies derivatives using the chain rule $\frac{d\tilde f}{d\lambda}=\frac{\partial\tilde f}{\partial\lambda_{\rm fast}}\frac{d\lambda_{\rm fast}}{d\lambda}+\frac{\partial\tilde f}{\partial\lambda_{\rm slow}}\frac{d\lambda_{\rm slow}}{d\lambda}$, and solves the resulting equations while treating $\lambda_{\rm fast}$ and $\lambda_{\rm slow}$ as independent variables. 

Here, I adopt $\tilde t:=\e t$ as my slow variable on the worldline $z^\mu$; its extension off the worldline will be discussed in the next section. As my fast variable I adopt $\phi_p$. Since the orbital radius and frequency evolve slowly, I write them as functions of $\tilde t$, $r_p(t,\e)=\tilde r_p(\e t, \e)$ and $\Omega(t,\e)=\tilde\Omega(\e t,\e)$, where   
\begin{align}
\tilde r_p(\tilde t,\e)&= \tilde r_0(\tilde t) + \e \tilde r_1(\tilde t) + \O(\e^2),\label{rp-expansion}\\
\tilde \Omega(\tilde t,\e) &=\tilde \Omega_0(\tilde t) + \e \tilde \Omega_1(\tilde t) + \O(\e^2).\label{Omega-expansion}
\end{align}
The orbital phase, $\phi_p(t,\e)=\tilde \phi_p(\e t,\e)$, is recovered from the frequency as
\beq\label{phip-expansion}
\tilde \phi_p(\tilde t,\e) = \frac{1}{\e}\int_0^{\tilde t}[\tilde \Omega_0(\tilde s) + \e \tilde \Omega_1(\tilde s)]d\tilde s + \O(\e).
\eeq
These expansions follow Ref.~\cite{Pound:10c}, itself inspired by Ref.~\cite{Hinderer-Flanagan:08}. However, unlike in those references, because the orbit is quasicircular, here the equation of motion contains no explicit dependence on the fast variable. Hence, the expansion considered in this section is not truly a multiscale one. Nevertheless, when combined with the expansion of the field in the next section, the use of the term ``multiscale'' becomes appropriate, and the use of $\phi_p$ as a fast variable becomes clear.

Because $\frac{D^2z^\mu}{d\tau^2} u_\mu=0$ (and because there is no motion in the $\theta$ direction), only two components of the equation of motion~\eqref{EOM} are independent. I choose to work with the $t$ and $r$ components. After substituting Eqs.~\eqref{rp-expansion}--\eqref{Omega-expansion} and using $d/dt=\e\, d/d\tilde t$, we find that the $t$ component reads
\beq
\e \frac{dU}{d\tilde t} = \frac{1}{U}[\e f_{1\rm self}^t+\e^2 f_{2\rm self}^t+\O(\e^3)],\label{EOM-t}
\eeq
and the $r$ component reads
\begin{align}
\e^2\frac{d^2r_p}{d\tilde t^2} +&\e^2 \frac{1}{U}\frac{dU}{d\tilde t}\frac{dr_p}{d\tilde t} -r_p\Omega^2 \nonumber\\
			&= -\frac{1}{r_p^2}+\frac{1}{U^2}[\e f_{1\rm self}^r+\e^2 f_{2\rm self}^r+\O(\e^3)].\label{EOM-r}
\end{align}
%and
%\begin{align}\label{EOM-phi}
%\e\frac{d\Omega}{d\bar t} +\e \frac{1}{U}\frac{dU}{d\bar t} \Omega &= \frac{1}{U^2}[\e f_{1\rm self}^\phi+\e^2 f_{2\rm self}^\phi+\O(\e^3)],
%\end{align}
%where $U:=\frac{dt}{d\tau}=1/\sqrt{1-r_p^2\Omega^2}$.

On the right-hand sides of these equation, we write the self-forces as functions of slow time, $f_{n\,\rm self}^\mu(\tilde t,\e)$, and expand them in powers of $\e$ at fixed $\tilde t$ to obtain
\beq\label{fself-expansion}
\e f_{1\,\rm self}^\mu+\e^2 f_{2\,\rm self}^\mu = \e\tilde f^\mu_1(\tilde t)+\e^2\tilde f^\mu_2(\tilde t)+\O(\e^2),
\eeq
where 
\begin{align}
\tilde f^\mu_1(\tilde t) &= f^\mu_{1\,\rm self}(\tilde t,0),\label{tildef1}\\ 
\tilde f^\mu_2(\tilde t) &= f^\mu_{2\,\rm self}(\tilde t,0)+\frac{\partial f^\mu_{1\,\rm self}}{\partial\e}(\tilde t,0).\label{tildef2}
\end{align}
Concrete expressions for $\tilde f_n^\mu$ are presented in Sec.~\ref{multiscale-summary}, but for the moment, these abstract ones suffice.

Substituting Eqs.~\eqref{rp-expansion}, \eqref{Omega-expansion}, and \eqref{fself-expansion} into the equations of motion~\eqref{EOM-t} and \eqref{EOM-r} leads straightforwardly to a sequence of equations for $\tilde r_n(\tilde t)$ and $\tilde \Omega_n(\tilde t)$. Specifically, the equations break down as follows: at $n$th order, the $r$ component of the equation of motion, Eq.~\eqref{EOM-r}, provides an equation for $\tilde \Omega_n$ as a function of $\tilde r_1,\ldots,\tilde  r_n$, and the $t$ component, Eq.~\eqref{EOM-t}, provides an evolution equation for $\tilde r_{n-1}(\tilde t)$.

%First, to account for the fact that $F^\mu$ is a functional of $x_p^\mu$, I write it as
%\beq
%F^\mu(t,\e) = \e F_1^\mu(r_p;r_p) + \e^2 F^\mu_2(r_p;r_p) +O(\e^3),
%\eeq
%where $F_n^\mu(r_p;r_p)$ are evaluated at $r_p$ and functionals of the orbit of radius $r_p$. In omitting any dependence on $t$ and $\phi_p$, I've foreshadowed the next section, which shows that these dependencies drop out due to the metric's (quasi)helical symmetry. Substituting the expansions~\eqref{rp-expansion} and \eqref{phip-expansion}, I rewrite the force as
%\beq\label{force-expansion}
%F^\mu(t,\e) = \e F_1^\mu(r_0;r_0) + \e^2 \hat F^\mu_2(r_0;r_0,r_1) +O(\e^3),
%\eeq
%where $\hat F^\mu_2:=F^\mu_2(r_0;r_0)+\left[\frac{\partial}{\partial r_0}F^\mu_1(r_0;r_0)+\frac{\delta}{\delta r_0}F^\mu_1(r_0;r_0)\right] r_1$, where $\frac{\partial}{\partial r_0}$ acts on the first argument in $F^\mu_1(r_0;r_0)$, and $\frac{\delta}{\delta r_0}$ on the second.

At zeroth order, only the $r$ component, Eq.~\eqref{EOM-r}, contains a nontrivial piece, from which we find
\beq\label{Omega0}
\tilde \Omega_0(\tilde t) = \sqrt{\frac{1}{\tilde r_0(\tilde t)^3}},
\eeq
a slowly evolving version of the frequency of the zeroth-order orbit in Sec.~\ref{near-zone}. 
%
%Beyond zeroth order, we require an expansion for $U$, which can be found from the normalization condition $U^2 g_{\mu\nu}\dot x_p^\mu \dot x_p^\nu=-1$. After involving Eq.~\eqref{Omega0}, we find
%\begin{align}\label{U-expansion}
%U^{-2} &= 1-\frac{3M}{r_0} - 2\e r_0^2\Omega_0\Omega_1+O(\e^3).
%\end{align}

Moving to linear order in $\e$, from the $t$ component, Eq.~\eqref{EOM-t}, we find an equation for the slow evolution of $\tilde r_0$:
\beq\label{r0}
\frac{d\tilde r_0}{d\tilde t} = -\frac{2\tilde r_0^2}{\tilde U_0^4} \tilde f_{1}^t,
\eeq
where $\tilde U_0=1/\sqrt{1-1/\tilde r_0}$.%here (and below), it is understood that $r_n$, $\Omega_n$, $U_0=1/\sqrt{1-1/r_0}$, and $\tilde f^\mu_n$ are all functions of $\tilde t$.

Staying at linear order, from Eq.~\eqref{EOM-r} we get an equation for $\tilde\Omega_1$ as a function of $\tilde r_0$ and $\tilde r_1$:
\beq\label{Omega1}
\tilde\Omega_1 = \frac{1}{2 \tilde r^{5/2}_0}\left[(1 - \tilde r_0)\tilde r^2_0\tilde f_{1}^r - 3 \tilde r_1\right].
\eeq
In the usual treatment of self-accelerated circular orbits (e.g., \cite{Pound:14c}), we expand the perturbed orbit at fixed frequency, setting $\tilde\Omega_1=0$ and thereby obtaining a formula for $\tilde r_1$ in terms of $\tilde f^r_1$. But because we account for dissipation here, all the quantities are evolving, and we no longer have the freedom to make that choice except at a single value of $\tilde t$, say $\tilde t=0$. That choice implies the initial condition
\beq\label{r1(0)}
\tilde r_1(0) = \frac{1}{3}[1 - \tilde r_0(0)]\tilde r^2_0(0)\tilde f_{1}^r(0).
\eeq
 
   %I will return to this issue below.

Finally, from the second-order term in Eq.~\eqref{EOM-t}, we get an evolution equation for $\tilde r_1(\tilde t)$: 
\begin{align}
\frac{d\tilde r_1}{d\tilde t} &=  -\frac{2\tilde r_0^2}{\tilde U_0^4}\bigg[\tilde f_2^t - (\tilde r_0-2)\tilde r_0  \tilde f_1^r \tilde f_1^t \nonumber\\
 							&\quad + 2 \frac{\tilde U_0^2}{\tilde r_0}\tilde r_1 \tilde f_1^t + \frac{1}{2}\tilde r_0\tilde U_0^2 \frac{d\tilde f_1^r}{d\tilde t}\bigg].\label{r1}
\end{align}
Although we can also solve the second-order piece of Eq.~\eqref{EOM-r}, doing so would only give us an ineffectual equation for $\tilde\Omega_2$ in terms of $\tilde r_2$; determining the evolution of $\tilde r_2$ would require carrying the expansion to third order.

Combining the results~\eqref{Omega0}--\eqref{r1} provides us with the first two terms in the expansion of the frequency~\eqref{Omega-expansion}. From the frequency, we can calculate the expansion~\eqref{phip-expansion} of the orbital phase $\phi_p$. The leading term in this expansion (``adiabatic order'' in the language of Ref.~\cite{Hinderer-Flanagan:08}) is constructed from Eqs.~\eqref{Omega0} and \eqref{r0}, and it requires only the first-order dissipative force $\tilde f^t_1$. The subleading term (``post-adiabatic order'') is constructed from Eqs.~\eqref{Omega1} and \eqref{r1} and requires the complete first-order force and the dissipative piece $\tilde f^t_2$ of the second-order force.\footnote{For an ordinary scalar-field model in flat spacetime, the conservative self-force $\tilde f^r_1$ would vanish. However, since we have leeway in choosing $\varphi^\res_1$ in the current model, we may choose it such that $\tilde f^r_1\not\equiv0$.} Computing those forces in a practical way necessitates combining the expansion of the equation of motion with a multiscale expansion of the field equations.

%   And from there we can calculate the gravitational wave phase.

\subsection{Expansion of the field}\label{multiscale-field}

%If we substitute the worldline \eqref{xp}, together with the expansions~\eqref{rp-expansion} and \eqref{Omega-expansion}, into the standard field equations of (the self-consistent version of) self-force theory, and if we then follow the usual rules for a two-timescale expansion, we obtain a first-order equation of the form

To expand the fields $\varphi_n$, we must choose slow and fast variables as fields on spacetime, not only on the worldline. I do so by appealing to the approximants~\eqref{phi1-uniform} and \eqref{S2-uniform} for the first-order field and second-order source. Far from the worldline, they are oscillatory functions of $\phi_p$, with amplitudes that slowly vary with retarded time $u$. Hence, as my slow and fast variables I adopt $\tilde u:=\e u$ and $\tilde \phi_p(\tilde u,\e)$. For conciseness, I write the latter as $\tilde \phi_p(\tilde u)$.

I write the coefficients in the harmonic expansion $\varphi_n = \sum \varphi_{n\ell m}Y_{\ell m}$ as $\varphi_{n\ell m}(t,r,\e)=\tilde\varphi_{n\ell m}(\e(t-r),r, \phi_p(t-r,\e),\e)$, where 
\begin{align}\label{phi-multiscale}
\tilde\varphi_{n\ell m}(\tilde u,r, \tilde\phi_p,\e) &= \tilde\varphi_{n\ell m}(\tilde u,r,\tilde\phi_p,0)\nonumber\\
														&\quad +\e\frac{\partial \tilde\varphi_{n\ell m}}{\partial\e}(\tilde u, r,\tilde\phi_p, 0)+\O(\e^2),
\end{align}
and I define the new field variables
\begin{align}
\tilde \varphi_{1\ell m}(\tilde u,r,\tilde\phi_p) &= \tilde\varphi_{1\ell m}(\tilde u,r,\tilde\phi_p,0),\\
\tilde \varphi_{2\ell m}(\tilde u,r,\tilde\phi_p) &= \tilde\varphi_{2\ell m}(\tilde u,r,\tilde\phi_p,0)\nonumber\\
														&\quad+\frac{\partial \tilde\varphi_{1\ell m}}{\partial\e}(\tilde u, r,\tilde\phi_p, 0).
\end{align}
The essential aspect of this expansion is that both $\tilde u$ and $\tilde\phi_p(\tilde u,\e)$ are held fixed while expanding in powers of $\e$. Each of the variables $\tilde \varphi_{n\ell m}$ I write in terms of $\tilde\phi_p$ as
\beq\label{multiscale-ansatz}
\tilde \varphi_{n\ell m}(\tilde u,r,\tilde\phi_p(\tilde u)) = \tilde R_{n\ell m}(\tilde u, r)e^{-im\tilde\phi_p(\tilde u)}.
\eeq
Even without the guidance of Eq.~\eqref{phi1-uniform}, factoring out $e^{-im\tilde\phi_p(\tilde u)}$ from the field is natural, given that $\rho$ is an oscillatory functional of $\phi_p(t)$ and that the solution propagates along null curves. The benefit of choosing an asymptotically null slow variable was also emphasized previously by Mino and Price~\cite{Mino-Price:08}, and it is made plain by the analysis in Sec.~\ref{near-zone}. 

Analogously, I write $\rho_{\ell m}(t,r,\e)=\tilde\rho_{\ell m}(\e(t-r),r, \phi_p(t-r,\e),\e)$ and define the new source variables
\begin{align}
\tilde \rho_{1\ell m}(\tilde u,r,\tilde\phi_p) &= \tilde\rho_{\ell m}(\tilde u,r,\tilde\phi_p,0),\\
\tilde \rho_{2\ell m}(\tilde u,r,\tilde\phi_p) &= \tilde\rho_{\ell m}(\tilde u,r,\tilde\phi_p,0)\nonumber\\
														&\quad+\frac{\partial \tilde\rho_{\ell m}}{\partial\e}(\tilde u, r,\tilde\phi_p, 0),
\end{align}
with 
\beq
\tilde \rho_{n\ell m}(\tilde u,r,\tilde\phi_p(\tilde u)) = \tilde \varrho_{n\ell m}(\tilde u, r)e^{-im\tilde\phi_p(\tilde u)}.
\eeq
We can find explict expressions for these quantities by substituting the expansions~\eqref{rp-expansion}--\eqref{phip-expansion} into Eq.~\eqref{rho} and then expanding functions of $\tilde t$ around $\tilde u=\tilde t - \e r$. Concretely, the latter expansions read $\tilde r_n(\tilde t) = \tilde r_n(\tilde u) +\e r \dot{\tilde r}_n(\tilde u)+\O(\e^2)$, $\tilde\Omega_n(\tilde t) = \tilde\Omega_n(\tilde u) + \e r\dot{\tilde\Omega}_n(\tilde u) +\O(\e^2)$, and 
\begin{align}
\tilde\phi_p(\tilde t) &= \tilde\phi_p(\tilde u)+r\tilde\Omega_0(\tilde u)+\e\!\left[\frac{1}{2}r^2\dot{\tilde\Omega}_0(\tilde u)+r\tilde\Omega_1(\tilde u)\right]\nonumber\\
						&\quad +\O(\e^2).
\end{align}
Note that $ \tilde\phi_p(\tilde u)\sim1/\e$ and that the delta function in $\rho$ enforces $r=r_p$, which keeps $\e r$ small. The final result is
\begin{align}
\tilde \varrho_{1\ell m}(\tilde u,r) &=N_{\ell m} \tilde U^{-1}_0(\tilde u)\frac{e^{-im\tilde\Omega_0(\tilde u)r}}{r^2}\delta(r\!-\!\tilde r_0(\tilde u)\!),\\
\tilde \varrho_{2\ell m}(\tilde u,r) &= -N_{\ell m}\tilde U^{-1}_0(\tilde u)\frac{e^{-im\tilde\Omega_0(\tilde u)r}}{r^2}\times\Bigg\{\nonumber\\
									&\quad \left[\tilde r_1(\tilde u)+r\dot{\tilde r}_0(\tilde u)\right]\!\delta'(r\!-\!\tilde r_0(\tilde u)\!)\nonumber\\
									&\quad +\tilde U^{-1}_0(\tilde u)\!\left[\tilde U_1(\tilde u)+r\dot{\tilde U}_0(\tilde u)\right]\!\delta(r\!-\!\tilde r_0(\tilde u)\!)\nonumber\\
									&\quad +im\!\left[r\tilde\Omega_1(\tilde u)+\frac{1}{2}r^2\dot{\tilde\Omega}_0(\tilde u)\right]\!\delta(r\!-\!\tilde r_0(\tilde u)\!)\!\Bigg\}.\label{varrho2}
\end{align}

We must also expand the nonlinear source term in Eq.~\eqref{phi2}. Using the method in Appendix~\ref{coupling formula}, we immediately find $S_{2}=\sum\tilde{S}_{2\ell m}(\tilde u, r)e^{-im\tilde\phi_p(\tilde u)}Y_{\ell m}+\O(\e)$, where $\tilde{S}_{2\ell m}(\tilde u, r)$ is given by Eq.~\eqref{calS2lm}. As in Sec.~\ref{near-zone}, for the purposes of this paper we are not interested in the contribution of $\Box\varphi_2^\P$ to this source, but given an explicit expression for $\varphi_2^\P$ (as we have in the gravity case), finding that contribution is straightforward; in line with the conditions on the puncture, $\varphi_n^\P$ possesses expansions analogous to Eqs.~\eqref{phi-multiscale} and \eqref{multiscale-ansatz}.

Finally, when substituting Eqs.~\eqref{phi-multiscale}--\eqref{multiscale} into the left-hand sides of Eqs.~\eqref{phi1}--\eqref{phi2}, and after writing the differential operators in terms of $(u,r)$ coordinates, we must be mindful that derivatives with respect to $u$ introduce additional factors of $\e$. Specifically, $\partial_u\tilde R_{n\ell m}=\e\partial_{\tilde u}\tilde R_{n\ell m}$ and
\beq
\partial_u e^{-im\tilde\phi_p(\tilde u)} = -im[\tilde\Omega_0(\tilde u)+\e\tilde\Omega_1(\tilde u)+\O(\e^2)]e^{-im\tilde\phi_p(\tilde u)}.
\eeq

After combining all of the above expansions in Eqs.~\eqref{phi1}--\eqref{phi2}, I group terms by powers of $\e$ \emph{at fixed $\tilde u$ and $\tilde\phi_p(\tilde u,\e)$}. This leads to the equations
\begin{align}
\partial_r^2 \tilde R_{n\ell m} &+\frac{2}{r}\left(1+im\tilde\Omega_0 r\right)\partial_r \tilde R_{n\ell m} \nonumber\\
 &+\frac{1}{r^2}\left[2im\tilde\Omega_0r-\ell(\ell+1)\right]\tilde R_{n\ell m} = \tilde{\cal S}_{n\ell m},\label{tildeRnlm-eqn}
\end{align}
where the sources are
\begin{align}
\tilde{\cal S}_{1\ell m} &= -4\pi \tilde\varrho_{1\ell m},\label{tildecalS1lm}\\
\tilde{\cal S}_{2\ell m} &= \tilde{S}_{2\ell m} -4\pi\tilde\varrho_{2\ell m}   \nonumber\\
						&\quad +2(\partial_{\tilde u}-im\tilde\Omega_1)\left(\partial_r\tilde R_{1\ell m}+\tfrac{1}{r}\tilde R_{1\ell m}\right).\label{tildecalS2lm}
\end{align}

Since we wish to match the solution in the near zone to a well-behaved solution in the far zone, we should not naively attempt to find the retarded solution to the above equations. Instead, we may follow the strategy described at the end of Sec.~\ref{near-zone}: choose a large-$r$ boundary at $r=\mathcal{R}$, cut off the retarded integrals at that point, and then add a homogeneous solution to account for the part of the source that lies at $r>{\cal R}$.

To ensure regularity at $r=0$, the added homogeneous solution must be regular there. In terms of the variables $\tilde R_{n\ell m}$, this implies
\beq\label{general-solution}
\tilde R_{n\ell m} = \tilde c^+_{n\ell m}(r)\tilde R^+_{\ell m}+[\tilde c^-_{n\ell m}(r)+k_{n\ell m}]\tilde R^-_{\ell m}, 
\eeq
with the known coefficients
\begin{align}
\tilde c^+_{n\ell m}(r) &= \int_0^r \frac{\tilde R^-_{\ell m}(r')\tilde{S}_{n\ell m}(r')}{\tilde W_{\ell m}(r')}dr',\label{c-}\\
\tilde c^-_{n\ell m}(r) &= \int_r^{\cal R} \frac{\tilde R^+_{\ell m}(r')\tilde{S}_{n\ell m}(r')}{\tilde W_{\ell m}(r')}dr',\label{c+}
\end{align}
and some unknown ($r$-independent) functions $k_{n\ell m}(\tilde u, \mathcal{R})$ that are to be determined by matching. This is the most general solution compatible with (i) the assumptions of the multiscale expansion and the ansatz~\eqref{multiscale-ansatz}, (ii) retarded propagation inside the near zone, and (iii) regularity at $r=0$.

For $m\neq0$ the homogeneous solutions are
\begin{align}
\tilde R^+_{\ell m} &= e^{-im\tilde\Omega_0 r}h^{(1)}_{\ell}(m\tilde\Omega_0 r),\label{tildeR+}\\
\tilde R^-_{\ell m} &= e^{-im\tilde\Omega_0 r}j_{\ell}(m\tilde\Omega_0 r),
\end{align}
and the Wronskian is
\beq\label{tildeWlm}
\tilde W_{\ell m} = -\frac{e^{-2im\tilde\Omega_0 r}}{im\tilde\Omega_0 r^2}.
\eeq
Note that the functional forms of Eqs.~\eqref{tildeR+}--\eqref{tildeWlm} differ from Eqs.~\eqref{Rlm}--\eqref{Wlm} not for any reason related to the multiscale expansion, but simply because of the change from $t$ to $u$ as the time coordinate. For $m=0$, the homogeneous solutions and Wronskian remain~\eqref{Rl0}--\eqref{Wl0}.

\subsubsection{First-order solution}
At first order, the integrals~\eqref{c-}--\eqref{c+} are independent of ${\cal R}$, and evaluating them gives us%the source is confined to the evolution zone, and so the integral~\eqref{c+} is independent of $\mathcal{R}$. Since $\mathcal{R}$ is arbitrary, the final result must be independent of it  from $$ can simplify matters by writing Eq.~ $\mathcal{R}$
\begin{align}
\tilde R_{1\ell m} &= \frac{4\pi i}{\tilde U_0}N_{\ell m}m\tilde\Omega_0j_\ell(m\tilde\Omega_0 \tilde r_<)h^{(1)}_\ell(m\tilde\Omega_0\tilde r_>)e^{-im\tilde\Omega_0r} \nonumber\\
							&\quad + k_{1\ell m}j_{\ell}(m\tilde\Omega_0 r)\label{tildeR1lm}
\end{align}
for $m\neq0$ and
\begin{align}\label{tildeR1l0}
\tilde R_{1\ell 0} = \frac{4\pi}{\tilde U_0}\frac{N_{\ell0}}{2\ell+1}\frac{\tilde r_<^\ell}{\tilde r_>^{\ell+1}} + k_{1\ell 0}r^\ell
\end{align}
for $m=0$. Here $\tilde r_{\lessgtr}={\rm min/max}(r,\tilde r_0(\tilde u))$.

Because the matching procedure is trivial at this order, I state the results immediately: comparison with the exact solution~\eqref{phi1-1/r-v2}, or its approximant~\eqref{phi1-uniform}, tells us
\beq\label{k1lm}
k_{1\ell m} = 0 \quad\text{for all $\ell m$}.
\eeq
This implies that unlike Eq.~\eqref{GWphi1-ret}, Eq.~\eqref{tildeR1lm} contains no oscillations with respect to $r$; the oscillatory behavior has been moved into the factor $e^{-im\tilde \phi_p(\tilde u)}$ in Eq.~\eqref{multiscale-ansatz}. At large $r$, the $m\neq0$ solution behaves as 
\beq\label{tildeR1lm-1/r}
\tilde R_{1\ell m} = \frac{4\pi (-i)^\ell N_{\ell m}j_\ell(m\tilde\Omega_0\tilde r_0)}{\tilde U_0r}+\O(1/r^2).
\eeq
When placed in Eq.~\eqref{phi-multiscale}, this uniformly approximates the exact solution~\eqref{phi1-1/r-v2} on intervals of retarded time $\lesssim1/\e$, as we can see by comparison with Eq.~\eqref{phi1-uniform}. Like the first-order Gralla-Wald expansion, it is also uniform in $r$ at fixed $u$, valid all the way to null infinity. But unlike the Gralla-Wald expansion, it behaves correctly at spatial infinity as well: the amplitudes $r\tilde R_{1\ell m}$ of the outgoing waves $\propto \frac{e^{-im\tilde\phi_p}}{r}$ decay to zero in the infinite past, $u\to-\infty$, and hence also at spatial infinity. 

In fact, it is likely that the approximation here is truly uniform, valid no matter how large the scaling of $u$, $t$, or $r$ with $\e$. To see this, note from Eq.~\eqref{phi1-uniform} that the domain of validity of $\tilde\varphi_1$ is entirely determined by how well we approximate the motion. And since there is no timescale longer than $1/\e$ in the motion, we likely have a uniform approximation to it on all scales.

\subsubsection{Second-order solution: introducing a puncture at infinity}\label{multiscale-2nd-order-field}
As in Sec.~\ref{near-zone}, to characterize the behavior of the second-order solution, I split it into two terms: $\tilde \varphi_{2} = \tilde\psi + \widetilde{\delta\varphi}_1$, with a corresponding split 
\beq\label{tildeR2lm-split}
\tilde R_{2\ell m} = \tilde R^{\tilde\psi}_{2\ell m} + \tilde R_{2\ell m}^{\widetilde{\delta\varphi}},
\eeq
where the first term is generated by $\tilde{S}_{2\ell m}$, and the second by the remaining terms in Eq.~\eqref{tildecalS2lm}. In analogy with $\GW{\psi}$ and $\delta\varphi_1$, $\tilde R^{\tilde\psi}_{2\ell m}$ is sourced by the explicitly nonlinear term in the field equation, while $\tilde R_{2\ell m}^{\widetilde{\delta\varphi}}$ is sourced by the expansion~\eqref{rp-expansion}--\eqref{Omega-expansion} and slow evolution of the worldline. For simplicity, I will only be interested in the large-$r$ and large-${\cal R}$ behavior of these fields. I will also place the unknown homogeneous solution $k_{2\ell m}\tilde R^-_{\ell m}$ entirely into $\tilde R^{\tilde\psi}_{2\ell m}$.

Begin with $\tilde R_{2\ell m}^{\widetilde{\delta\varphi}}$. It is not hard to see from Eqs.~\eqref{general-solution}--\eqref{c-} and \eqref{S2lm} that at leading order in $1/r$, the solution behaves as an outgoing wave of the form 
\beq
\tilde R_{2\ell m}^{\widetilde{\delta\varphi}}e^{-im\tilde\phi_p(\tilde u)}\sim\frac{e^{-im\tilde\phi_p(\tilde u)}}{r}. 
\eeq
The piece of $\tilde R_{2\ell m}^{\widetilde{\delta\varphi}}$ sourced by $\delta$ and $\delta'$ terms is immediately found to have this behavior. The piece sourced by the $\tilde\Omega_1$ and $\partial_{\tilde u}\tilde R_{1\ell m}$ terms in Eq.~\eqref{S2lm} requires only slghtly more examination. From the approximation
\begin{align}
\left(\partial_r+\tfrac{1}{r}\right)\left[h^{(1)}_\ell(m\tilde\Omega_0 r)e^{-im\tilde\Omega_0 r}\right] &= -\frac{(-i)^\ell \ell(\ell+1)}{2m^2\tilde\Omega_0^2r^3}\nonumber\\
																																	&\quad +\O(1/r^4),\label{falloff}
\end{align}
we find that this part of the source falls off as $1/r^3$. A brief analysis then establishes the desired $\sim 1/r$ falloff of the solution. Therefore, we can conclude that the multiscale expansion has eliminated the unwanted secular growth discussed in Sec.~\ref{2nd-order-dphi}. Furthermore, we can easily show from Eqs.~\eqref{falloff} and~\eqref{general-solution}--\eqref{c-} that the ${\cal R}$ dependence of $\tilde R_{2\ell m}^{\widetilde{\delta\varphi}}$ falls to zero as ${\cal R}\to\infty$. Since $\tilde R_{2\ell m}$ must ultimately be independent of ${\cal R}$, this decaying ${\cal R}$ dependence must be exactly countered by an opposite dependence in $k_{2\ell m}$. Hence, we can simply send ${\cal R}\to\infty$ in the integrals in $\tilde R_{2\ell m}^{\widetilde{\delta\varphi}}$ and absorb the change into $k_{2\ell m}$. 

However, note that our choice of slow time variable is essential for these desirable falloff properties. If we had used $\tilde t$ instead of $\tilde u$ as our slow variable away from the worldline, the desirable properties would decidedly not manifest: one can easily check that the source term analogous to the final line of Eq.~\eqref{tildecalS2lm} would go as $\sim\partial_{\tilde t}\frac{e^{im\tilde\Omega_0(\tilde t)r}}{r}\sim r^0$, leading to poor behavior at large $r$ (and ${\cal R}$). Such behavior in the large-$r$ region of the near zone would not be catastrophic, as we ultimately must match to a far-zone solution in any case. Indeed, the bad behavior we would find would precisely correspond to the growth in $r$ that arises from expanding the exact solution~\eqref{phi1-1/r-v2} in powers of $\e$ at fixed $\tilde t$ instead of at fixed $\tilde u$. However, the nicer behavior we obtain using $\tilde u$ simplifies the matching procedure by improving the uniformity of the multiscale expansion, bringing it closer to the exact behavior in the far zone.  

Now turn to the second field in Eq.~\eqref{tildeR2lm-split}, $\tilde R^{\tilde\psi}_{2\ell m}$. Here the analysis of Eqs.~\eqref{general-solution}--\eqref{c+} closely follows Sec.~\ref{2nd-order-psi}, with obvious alterations as necessary. Equations~\eqref{calS2lm} and \eqref{tildeR1lm-1/r} show that the analogue of Eq.~\eqref{S2lm-1/r2} reads
\beq
\tilde S_{2\ell m} = \frac{\tilde S^{(-2)}_{2\ell m}(\tilde u)}{r^2}+\O(1/r^3).
\eeq
Unlike in Eq.~\eqref{S2lm-1/r2}, the $1/r^2$ term here contains no oscillatory factor, a consequence of  Eq.~\eqref{tildeR1lm-1/r}. But because of the oscillatory factors in $\tilde R^\pm_{\ell m}$ and $\tilde W_{\ell m}$, the integrals~\eqref{c-}--\eqref{c+} are identical to those in Sec.~\ref{S2 harmonic decomposition} (except for the cutoff at $r={\cal R}$).

For $\ell>0$, we can easily show that, as with $\tilde R_{2\ell m}^{\widetilde{\delta\varphi}}$, we can extend the integrals to ${\cal R}\to\infty$ and absorb the difference into $k_{2\ell m}$. We then find the analogues of Eqs.~\eqref{R-22lmneq0} and \eqref{R-22lneq00} are
\beq\label{tildeR-22lm}
\tilde R^{(-2)}_{2\ell m} = \frac{(\tilde C_{\ell m}+\tilde S^{(-2)}_{2\ell m}\ln r)}{2im\tilde{\Omega}_0r}+k'_{2\ell m}j_{\ell}(m\tilde\Omega_0r)+\O(r^{-2}\ln r)
\eeq
for the $m\neq0$ modes, and
\beq\label{tildeR-22lm=0}
\tilde R^{(-2)}_{2\ell 0} = -\frac{\tilde S^{(-2)}_{2\ell0}}{\ell(\ell+1)} +k'_{2\ell 0}r^{\ell} + \O(1/r^{\ell+1})
\eeq
for the $m=0$, $\ell>0$ modes. Here $k'_{2\ell m}(\tilde u)=k_{2\ell m}(\tilde u,\infty)$ is independent of ${\cal R}$.

For $\ell=0$, we find
\begin{align}
\tilde R^{(-2)}_{200} &=  \left(\frac{r^+}{r}-1+\ln\frac{r}{{\cal R}}\right)\tilde S^{(-2)}_{200}+k_{200}.\label{tildeR-2200}
\end{align}
We may hence write the total monopole mode $\tilde R_{200}$ as
\begin{align}
\tilde R_{200} &=  \ln(r) \tilde S^{(-2)}_{200}+k'_{200}+\O(r^{-1}\ln r),\label{tildeR-2200-v2}
\end{align}
where $k'_{200}(\tilde u)=k_{200}(\tilde u,{\cal R})+\ln({\cal R})\tilde S^{(-2)}_{200}$ is independent of ${\cal R}$. Note that if we had not restricted our expansion to the near zone---that is, if we had set $k_{2\ell m }=0$ and let ${\cal R}\to\infty$---then we would have encountered precisely the same logarithmic divergence in Eq.~\eqref{tildeR-2200} as we did in the Gralla-Wald expansion. Unlike in the Gralla-Wald case, here this behavior is  not caused by a nonuniform approximation to $\varphi_1$: the approximation $\tilde\varphi_1$ is uniform in space, behaving correctly both at null and spatial infinity, and it produces a well-behaved source term $\tilde S^{(-2)}_{200}$ that decays  to zero at spatial infinity. Instead, the divergence is caused by a failure of the hypotheses of the multiscale expansion. I will discuss this further in Sec.~\ref{matching}. However, in the near zone the expansion is well behaved through second order, which is all we require to compute $\tilde f^t_2$; this in turn is all we require for the post-adiabatic approximation described in Sec.~\ref{multiscale-motion}. 

Allow me to summarize and describe how the above allows us to write the general solution~\eqref{general-solution} in a more convenient form. For all modes $\ell>0$, we can set ${\cal R}=\infty$, leaving the matching procedure to determine the ${\cal R}$-independent constants $k'_{2\ell m}=k_{2\ell m}({\cal R}=\infty)$. To streamline the discussion in Sec.~\ref{multiscale-summary}, I state in advance that matching will show 
\beq\label{k2l>0}
k'_{2\ell m} = 0 \quad\text{for }\ell\neq0,
\eeq
as one might expect from the fact that only the $\ell=0$ solution was ill defined in Sec.~\ref{near-zone}. In other words, for $\ell\neq0$  we need not have restricted the solution to the near zone. (Note this is true only because I used $\tilde u$, not $\tilde t$, as my slow time.) 

For the $\ell=0$ mode, this is not the case, but  we can nevertheless find a more convenient form of the general solution~\eqref{general-solution}. I achieve this using the same method as was used to deal with the behavior near the particle: introducing a puncture. Define the \emph{puncture at infinity}
\beq\label{phi-infty}
\tilde \varphi^{\infty}(\tilde u,r) = \theta(r-r^{\infty})\ln(r) \tilde S^{(-2)}_{200}(\tilde u),
\eeq
where $r^\infty(\tilde u)>r^+(\tilde u)$ is arbitrary. Then we can define an effective variable $\tilde R^{\rm eff}_{200}:=\tilde R_{200}-\tilde \varphi^{\infty}-k'_{200}$ and transfer $\tilde \varphi^{\infty}$ to the right-hand side of the field equation~\eqref{tildeRnlm-eqn}, leading to the equation $(\partial^2_r+2r^{-1}\partial_r)\tilde R^{\rm eff}_{200}=\tilde{\cal S}^{\rm eff}_{200}$, with the effective source 
\begin{align}
\tilde{\cal S}^{\rm eff}_{200} &= \tilde{\cal S}_{200} - (\partial^2_r+2r^{-1}\partial_r)\tilde \varphi^{\infty}\\
										&=  \tilde{\cal S}_{200} - \frac{\tilde S^{(-2)}_{200}}{r^2} \quad \text{for }r>r^\infty.
\end{align}
Because $k'_{200}$ is a homogeneous solution (at fixed $\tilde u$), it does not appear in the equation for $\tilde R^{\rm eff}_{200}$. The source falls off as $1/r^3$, and we can write the solution as
\beq\label{effective-solution}
\tilde R^{\rm eff}_{200} = \tilde c^{\rm eff+}_{200}\tilde R^+_{\ell m}+\tilde c^{\rm eff-}_{200}\tilde R^-_{\ell m}, 
\eeq
where
\begin{align}
\tilde c^{\rm eff+}_{200} &= \int_0^r \frac{\tilde R^-_{00}(r')\tilde{\cal S}^{\rm eff}_{200}(r')}{\tilde W_{00}(r')}dr',\\
\tilde c^{\rm eff-}_{200} &= \int_r^\infty \frac{\tilde R^+_{00}(r')\tilde{\cal S}^{\rm eff}_{200}(r')}{\tilde W_{00}(r')}dr'.
\end{align}
The physical field can then be recovered using
\beq\label{physical-solution}
\tilde R_{200} = \tilde R^{\rm eff}_{200} + \tilde \varphi^{\infty} + k'_{200}.
\eeq

With this setup, one can solve all the field equations on an infinite range of $r$ without worrying about any homogeneous solutions. Taking Eqs.~\eqref{k1lm} and \eqref{k2l>0} for granted, this yields the correct physical solution up to the single function $k'_{200}(\tilde u)$, which (once determined by matching) can be added as a final step.

\subsection{Combining the expansions of the worldline and of the field}\label{multiscale-summary}
In the preceding two sections, the expansions of the fields $\varphi_n$ and worldline $z^\mu$ left the fields as functionals of the worldline variables and the worldline variables as functionals of the fields. The two  are linked by the expansion~\eqref{fself-expansion} of the self-force. Substituting the expansions~\eqref{rp-expansion}, \eqref{Omega-expansion}, \eqref{phi-multiscale}, and \eqref{multiscale-ansatz} into the force~\eqref{fnself}, we find that the relevant components of the first-order self-force~\eqref{tildef1} are given by
\begin{align}
\tilde f^t_1(\tilde t) &= \sum_{\ell m}im\tilde\Omega_0N_{\ell m}\tilde R^{\res}_{1\ell m},\label{f1t}\\
\tilde f^r_1(\tilde t) &= \sum_{\ell m}N_{\ell m}\left[\partial_{\tilde r_0}\tilde R^{\res}_{1\ell m}+im\tilde\Omega_0\tilde R^{\res}_{1\ell m}\right],\label{f1r}
\end{align}
and those of the second-order self-force~\eqref{tildef2} are given by
\begin{align}
\tilde f^t_2(\tilde t) &= \sum_{\ell m}N_{\ell m}\Bigg\{ im\tilde\Omega_0\tilde R^{\res}_{2\ell m} + im\tilde\Omega_1\tilde R^{\res}_{1\ell m}\nonumber\\
							&\quad +im\tilde\Omega_0\tilde r_1\partial_{\tilde r_0}\tilde R^{\res}_{1\ell m}-\partial_{\tilde t}\tilde R^{\res}_{1\ell m} 
							+ \tilde U_0^2\bigg[\partial_{\tilde t}\tilde R^{\res}_{1\ell m}\nonumber\\
							&\quad +\frac{d\tilde r_0}{d\tilde t}\left(\partial_{\tilde r_0}\tilde R^{\res}_{1\ell m}+im\tilde\Omega_0\tilde R^{\res}_{1\ell m}\right)\bigg]\Bigg\},\label{f2t}\\
\tilde f^r_2(\tilde t) &= \sum_{\ell m}N_{\ell m}\Big[\partial_{\tilde r_0}\tilde R^{\res}_{2\ell m}+im\tilde\Omega_0\tilde R^{\res}_{2\ell m} \nonumber\\
							&\quad -\partial_{\tilde t}\tilde R^{\res}_{1\ell m}+\tilde r_1\partial_{\tilde r_0}^2\tilde R^{\res}_{1\ell m}
							+im\tilde\Omega_1\tilde R^{\res}_{1\ell m}\Big].\label{f2r}
\end{align}
(I include $\tilde f^r_2$ for completeness, but as described in Sec.~\ref{multiscale-motion}, it does not appear in the equations of motion at the level of long-term accuracy considered here.) In these expressions, $\tilde\Omega_n$ and $\tilde r_n$ are evaluated at $\tilde t$, and $\tilde R^{\res}_{n\ell m}(\tilde u,r)=\tilde R_{n\ell m}(\tilde u,r)-\tilde R^\P_{n\ell m}(\tilde u,r)$ and its derivatives are evaluated at $(\tilde u,r)=(\tilde t, \tilde r_0(\tilde t)\!)$; $\partial_{\tilde t}$ acts on the first argument, $\partial_{\tilde r_0}$ on the second. Because the dependence on $\tilde u$ in $\tilde R^{\res}_{n\ell m}(\tilde u,r)$ comes in the form of a dependence on $\tilde r_0(\tilde u)$, we may also rewrite $\partial_{\tilde t}$ in terms of an $\tilde r_0$ derivative. 

With these results, we can combine the expansions of the fields and worldline in the following practical prescription:%. First use Eqs.~\eqref{Omega0} and \eqref{Omega1} to eliminate $\Omega_n$ in favor of $r_n$ in the field equations~\eqref{tildeRnlm-eqn} and in the forces~\eqref{f1t}--\eqref{f2r}, and use Eq.~\eqref{r0dot} we have the following prescription: 
\begin{enumerate}
%\item Solve Eq.~\eqref{Omega0} for the zeroth-order frequency $\Omega_0$ as a function of the zeroth-order orbital radius $r_0$
\item Eliminate $\tilde\Omega_0$ in favor of $\tilde r_0$ in Eqs.~\eqref{tildeRnlm-eqn} and~\eqref{tildecalS1lm} using Eq.~\eqref{Omega0}. Solve Eq.~\eqref{tildeRnlm-eqn} for the first-order field $\tilde R_{1\ell m}$ as a function of $r$ and $\tilde r_0$ using Eq.~\eqref{general-solution} (with $k_{1\ell m}=0$). This yields Eqs.~\eqref{tildeR1lm}--\eqref{k1lm}, with $\tilde\Omega_0$ given by Eq.~\eqref{Omega0}.
\item Construct the first-order self-force $\tilde f^\mu_1$ as a function of $\tilde r_0$ using Eq.~\eqref{f1t}--\eqref{f1r}.
\item Eliminate $\tilde\Omega_1$ and $\frac{d\tilde r_0}{d\tilde t}$ in favor of $\tilde r_0$ and $\tilde r_1$ in Eqs.~\eqref{tildeRnlm-eqn} and~\eqref{tildecalS2lm} using Eqs.~\eqref{Omega1} and \eqref{r0}. Solve Eq.~\eqref{tildeRnlm-eqn} for the second-order field $\tilde R_{2\ell m}$ as a function of $\tilde r$, $\tilde r_0$, and $\tilde r_1$. For $\ell>0$, use Eq.~\eqref{general-solution} with ${\cal R}=\infty$ and $k_{2\ell m}=0$. For $\ell=0$, use Eqs.~\eqref{effective-solution} and \eqref{physical-solution}.
\item Construct the second-order self-force $\tilde f^t_2$ as a function of $\tilde r_0$ and $\tilde r_1$ using Eq.~\eqref{f2t}. 
%\item Solve Eq.~\eqref{r0dot} for the evolution of the orbital radius $r_0(\tilde t)$
\item Choose an initial value of $\tilde r_0$ and solve Eq.~\eqref{r0} for $\tilde r_0(\tilde t)$.
\item Choose Eq.~\eqref{r1(0)} as an initial value of $\tilde r_1$. Solve Eq.~\eqref{r1} for $\tilde r_1(\tilde t)$.
\item Construct the evolution of the orbital frequency $\tilde\Omega_0(\tilde t)+\e\tilde\Omega_1(\tilde t)$ from $\tilde r_0(\tilde t)$ and $\tilde r_1(\tilde t)$ using Eqs.~\eqref{Omega0} and \eqref{Omega1}.
%\item Solve Eq.~\eqref{R2} for the second-order field $\tilde R_{2\ell m}(r)$, and from it construct  the second-order force $\tilde f^\mu_2$ using Eqs.~\eqref{f2t}--\eqref{f2r}
%\item Solve Eq.~\eqref{r1} for the evolution of the radius's correction $r_1(\tilde t)$
%\item Compute the orbital frequency $\Omega_0(\tilde t)+\e\Omega_1(\tilde t)$ from $r_0(\tilde t)$ and $r_1(\tilde t)$, and from it 
\item Compute the phase evolution $\tilde\phi_p(\tilde t)$ using Eq.~\eqref{phip-expansion}.
\item Construct the time-domain $\ell m$ modes $\tilde\varphi_{n\ell m}$ from Eq.~\eqref{multiscale-ansatz}.
\end{enumerate}
One may note that in this toy model, one need not even know $k'_{200}$, as it has no effect on the approximation; it does not contribute to the force~\eqref{f2t}. This is a consequence of the invariance of the model under the gauge transformation $\varphi_n\to\varphi_n+\text{constant}$. I discuss the relevance of this to the gravity case in Sec.~\ref{lessons}.

The procedure outlined above is in the same spirit as the one used in first-order gravity in Ref.~\cite{Warburton-etal:12}: first compute the self-force as a function of orbital parameters, in this case $\tilde r_0$ and $\tilde r_1$, and then evolve an orbit through the parameter space as a post-processing step. In practice, this allows one to quickly evolve many orbits once one has populated the parameter space with self-force values. However, if one wished to avoid populating the parameter space in advance, one could straightforwardly reorganize the above prescription to evolve a single orbit, simply solving for the fields $\tilde R_{n\ell m}$ and parameters $\tilde r_n$ at each time step $\tilde u$.

Either way, the scheme developed here yields a uniformly first-order--accurate approximation to the total field $\e\varphi_1+\e^2\varphi_2$ on the radiation reaction time and all the way to $r\to\infty$. But one should note that it does \emph{not} provide a uniformly \emph{second}-order--accurate approximation. The omitted terms of order $\e$ in Eq.~\eqref{phip-expansion} lead to $\O(\e)$ errors in $\tilde\varphi_{1}$ through Eq.~\eqref{multiscale-ansatz}, and these errors are of the same size as $\e^2\tilde\varphi_2$. Obtaining a uniformly second-order--accurate approximation on the radiation-reaction time would require a third-order approximation to the equation of motion. In the next section, I will also describe the second-order approximation's lack of uniformity in $r$. However, these limitations do not affect second-order quantities constructed on the orbital timescale, such as the Detweiler redshift discussed in Ref.~\cite{Pound:14c}. Nor do they seriously hamper gravitational waveform generation: the primary goal in waveform modeling is to track the \emph{phase} accurately over the complete inspiral. For the phase, we require the second-order field only through its appearance in the second-order equation of motion; for the amplitude, first-order accuracy should suffice. %To accurately track the phase, all we require from the second-order field is its contribution to the second-order dissipative force, which the scheme presented here provides.

%%%%%%%%%%%%%%%%%%%%%%%%%%%%%%%%%%%%%%%%%%%%%%%%%%%%%%%
\section{Matching to the exact solution in the far zone}\label{matching}

%By now the reader may have been struck with the conviction that forcing a frequency-domain analysis on this problem is wrongheaded. It imposes a periodicity that simply does not exist. 
In this section, I show how to fix the functions $k'_{2\ell m}(\tilde u)$ in the multiscale expansion by matching to the exact solution in the far zone. The procedure confirms that $k'_{2\ell m}=0$ for all $\ell>0$, and it uniquely determines $k'_{200}$. Crucially, only minimal information from the exact solution is required to achieve this.

%To achieve this, I construct the general form of the retarded solution in the far zone. %In the buffer region $1/\e^{p>1}\gg r\gg 1$ that lies between the evolution zone and the  far zone, the multiscale expansion must agree with this time-domain solution. Put another way, if the far-zone expansion is re-expanded for small $r$, it must agree with the re-expansion of the evolution-zone expansion for large $r$.

%\beq
%\varphi_1 =  \int\frac{\rho(t-|\vec{x}-\vec{x}'|,\vec{x}')}{|\vec{x}-\vec{x}'|}d^3x'
%\eeq  
%      
%\beq
%\varphi_2 =  -\frac{1}{4\pi}\int\frac{S(t-|\vec{x}-\vec{x}'|,\vec{x}')}{|\vec{x}-\vec{x}'|}d^3x'
%\eeq

%My emphasis in this section is on developing a scheme that requires no direct access to the retarded integrals~\eqref{} and \eqref{}. As in post-Newtonian theory, the goal is to obtain a general solution in the near zone, a general solution in the far zone, and then fix both solutions via matching. 

\subsection{The general solution of Blanchet and Damour}
In a series of classic papers~\cite{Blanchet-Damour:86,Blanchet:87,Blanchet-Damour:88, Blanchet-Damour:89,Blanchet-Damour:92}, Blanchet and Damour derived the general form of the retarded solution to the PM field equations at all points outside any material source. They also developed an algorithm for building a global solution by matching this general form to an expansion in a suitable, smaller zone containing the matter. Although their methods were not designed for my specific toy model, they can be applied directly to it. In particular, Ref.~\cite{Blanchet-Damour:88}, which I hereafter refer to as BD, contains much of the relevant analysis for our current problem, and I adhere to it as much as possible.

The general form of a retarded solution is as follows. At all points $r>r_p(u)$, the solution to Eq.~\eqref{phi1} reads
\beq\label{phi1-BD}
\varphi_1 = \sum_\ell \frac{(-1)^\ell}{\ell!}\partial_L\frac{F^1_L(u)}{r},
\eeq
where the partial derivatives are taken at fixed $t$ and hence act on the dependence on $u$. This is the generic form of a homogeneous solution containing no incoming waves. In the present context, the set of functions $F^1_L(u)$ are to be determined by matching to the expansion in the evolution zone. Again at points $r>r_p(u)$, the  retarded solution to Eq.~\eqref{phi2} reads
\beq\label{phi2-BD}
\varphi_2 = \varphi^{\rm part}_2 + \varphi_2^{\rm hom},
\eeq
where
\beq\label{phi2-hom}
\varphi_2^{\rm hom} = \sum_\ell \frac{(-1)^\ell}{\ell!}\partial_L\frac{F^2_L(u)}{r}
\eeq
is another homogeneous solution containing no incoming radiation, and
\beq\label{phi2-part}
\varphi^{\rm part}_2 = {\rm FP}\Box^{-1}_{\rm ret}(r^B S[F^1_L])
\eeq
is a particular solution also containing no incoming radiation. To describe the right-hand side of Eq.~\eqref{phi2-part}, it will be convenient to limit the construction to the points $r>r^+(u)$, where the puncture field vanishes. In this case, $S[F^1_L]$ is the source $t^{\alpha\beta}\nabla_{\alpha}\varphi_1\nabla_\beta \varphi_1$ with $\varphi_1$ given by  Eq.~\eqref{phi1} \emph{for all values of $r>0$}. (I omit a sub- or superscript 2 on $S$ to keep the notation compact.) $\Box^{-1}_{\rm ret}$ denotes integration against the standard retarded Green's function over all spacetime, and ``FP'' denotes the ``finite part'' (i.e., the coefficient of $B^0$) in the Laurent series around $B=0$.\footnote{The FP operation used in BD is slightly more complicated and makes large-$r$ behavior easier to deal with. I avoid that construction to forestall any suspicion that Eq.~\eqref{phi2-BD} involves regularization at infinity.} I stress that this is not a regularization procedure but simply a (rigorous) way of finding a particular solution to Eq.~\eqref{phi2} at each point $r>r^+(u)$.\footnote{To see this, note that Eq.~\eqref{phi2-part} can be written as the sum of (i) the retarded integral of the true source over the region $r>r^+(u)$ and (ii) a homogeneous solution given by the finite part of the retarded integral of the fictitious source $r^B S[F^1_L]$ over the region $r<r^+(u)$. One could instead take a particular solution $\varphi^{\rm part}_2$ from the PM methods of Will et al., which do not involve the FP operation. For example, Eq.~(6.105) of Ref.~\cite{Poisson-Will:14} would provide such a solution.} By construction, the particular solution is a retarded field, containing no incoming radiation, but it is not the retarded solution to the real problem; an additional homogeneous solution~\eqref{phi2-hom} is required to complete the solution. The advantage of this construction is that it allows us to build a particular solution at points $r>r^+(u)$ without any knowledge of the behavior of $\varphi_1$ at points $r<r^+(u)$. That information about $\varphi_1$ determines the homogeneous solution~\eqref{phi2-hom}, the terms in which, like those in Eq.~\eqref{phi1-BD}, are to be found by matching.% to the multiscale expansion.

Given the form~\eqref{phi1-BD} for $\varphi_1$, we may write the source $S$  as
\beq\label{S=SLnL}
S = \sum_{\ell}\sum_{k\geq2} \frac{1}{r^k}S^{(-k)}_L(u)\nhat^L.\vspace{0pt}
\eeq
Expressing each $S^{(-k)}_L(u)$ in terms of the set of functions $\{F^1_L\}_\ell$ is straightforward but laborious; doing so involves simple applications of identities in Appendix A of Ref.~\cite{Blanchet-Damour:86}. However, here we do not explicitly require such expressions, since I eventually convert to coefficients of ordinary scalar harmonics using 
\begin{align}
F^n_L\nhat^L &= \sum_m F^n_{\ell m}Y_{\ell m},\label{FLnL=FlmYlm}\\
S^{(-k)}_L\nhat^L &= \sum_m S^{(-k)}_{2\ell m}Y_{\ell m},\label{SLnL=SlmYlm}
\end{align}
allowing me to use the coupling formulas in Appendix~\ref{coupling formula} of this paper. (I reinsert the subscript 2 on the source at the level of $\ell m$ modes to make the clearest link to previous sections; $S^{(-2)}_{2\ell m}(u)$ is precisely the function approximated by $\tilde S^{(-2)}_{2\ell m}(\tilde u)$, for example.)
%and
%\beq
%S_{\ell m}(u,r) = \sum_{k=2}^{2\ell+3}\frac{1}{r^k}S_{\ell m k}(u).
%\eeq

As described in BD, for each term in the source \eqref{S=SLnL}, the retarded integral appearing in Eq.~\eqref{phi2-part} can be simplified to\begin{widetext}
\beq\label{ret-klm}
\Box^{-1}_{\rm ret}\left(r^{B-k} S^{(-k)}_L\nhat^L\right) = \frac{1}{K(B,k)} \int_r^\infty dz\, S^{(-k)}_L(t-z)\hat\partial_L\left[\frac{(z-r)^{B-k+\ell+2}-(z+r)^{B-k+\ell+2}}{r}\right],
\eeq\end{widetext}
where 
\beq
K(B,k) = 2^{B-k+3}\frac{(B-k+2)!}{(B-k-\ell+1)!}.%(B-k+2)(B-k+1)\cdots(B-k-\ell+2).
\eeq

Only the most slowly falling term in the source, $r^{-2}S^{(-2)}_L\nhat^L$, will be of interest to us here; as mentioned previously, terms  that fall off faster than $1/r^2$ generate  retarded solutions that fall off as $\sim 1/r$. The matching procedure for those solutions would yield only trivial results. So in what follows, I will specialize Eq.~\eqref{ret-klm} to $k=2$. Because $1/r^2$ is integrable at $r=0$, for this term in the source we can replace the FP operation with $\lim_{B\to0}$. We could even avoid the limit and simply begin with the retarded integral of $r^{-2} S^{(-k)}_L\nhat^L$. However, it is advantageous to work with the general expression~\eqref{ret-klm} and take the finite part at a convenient stage. %BD shows that for $k=2$, the FP operation is equivalent to simply taking the limit $B\to0$, but it is convenient to take that limit only at the end of the calculation. 

\subsection{Matching at first order}
In the present context, the first-order matching procedure may be bypassed entirely. We could simply assume an implicit functional relationship $F^1_{\ell m}(u)=\tilde F^1_{\ell m}[\tilde\varphi_{1\ell m}]+\O(\e)$ and then construct $S^{(-2)}_{\ell m}$ from $\tilde\varphi_{1\ell m}$. However, for completeness, I go through the process in some detail. In Sec.~\eqref{multiscale}, we found asymptotic expressions,~\eqref{tildeR1l0} and~\eqref{tildeR1lm-1/r}, for $\varphi_1$ by first performing a multiscale expansion and then performing a large-$r$ expansion. The matching condition states that we must obtain the same results when we reverse these operations, first performing the large-$r$ expansion and then the multiscale expansion. (Note that this condition immediately implies the result~\eqref{k1lm}, which I will take for granted below.) 

I first expand Eq.~\eqref{phi1-BD} for large $r$. Following the manipulations above Eq.~\eqref{phi1-1/r}, this yields 
\beq\label{varphi1F-1/r}
\varphi_1 = \frac{1}{r}\sum_{\ell m}\frac{Y_{\ell m}}{\ell!}\frac{d^\ell F^1_{\ell m}}{du^\ell} +\O(1/r^2),
\eeq
where I have used Eq.~\eqref{FLnL=FlmYlm}. 

To perform the matching, I expand $F^1_{\ell m}(u,\e)$ as $\tilde F^1_{\ell m}(\tilde u,\tilde\phi_p(\tilde u))+\O(\e)$. We then have 
\beq
\varphi_1 = \frac{1}{r}\sum_{\ell m}\frac{Y_{\ell m}}{\ell!}\tilde \Omega_0^\ell \frac{\partial^\ell}{\partial\tilde\phi_p^\ell}\tilde F^1_{\ell m}+\O(\e,1/r^2),
\eeq
and comparison with Eqs.~\eqref{tildeR1l0} and~\eqref{tildeR1lm-1/r} establishes
\begin{align}
\tilde F^1_{00} &= \frac{4\pi}{\tilde U_0},\\
\tilde F^1_{\ell m} &= \frac{4\pi \ell!\, N_{\ell m}}{(m \tilde\Omega_0)^{\ell}\tilde U_0} j_\ell(m\tilde\Omega_0\tilde r_0)e^{-im\tilde\phi_p(\tilde u)}.
\end{align}
One would have to carry on to higher orders in $1/r$ to obtain results for $\tilde F^1_{\ell 0}$ with $\ell>0$. However, as we already know, only the $m\neq0$ modes of $\varphi_1$ are required to calculate $S^{(-2)}_{2\ell m}$. 

As previously stated, at first order the multiscale approximation is well behaved out to null and spatial infinity. Although this makes the matching procedure trivial at first order, the procedure does illuminate the reason for the uniform accuracy: the general retarded solution~\eqref{phi1-BD} cleanly separates into $u$ and $r$ dependence. The functional dependence on the worldline is entirely contained in the dependence on $u$, and an expansion in powers of $\e$ at fixed $(\tilde u,\tilde\phi_p)$ only involves the functions $F_L^1(u)$; the expansion does not interact with $\varphi_1$'s $r$ dependence, and so it places no restriction on the magnitude of $r$. Similarly, how far we can get toward spatial infinity depends only on $F_L^1(u)$; if the approximation is accurate over a range of time $u$ along null infinity, then it is accurate over that same range of $r$ at fixed $t$. [In the case of the Gralla-Wald expansion, the same reasoning applies at fixed $u$, but at fixed $t$ the expansion fails because any dependence on $\e u=\e(t-r)$ in $F^1_L(u)$ is expanded in powers of $\e r$, restricting the range of $r$.] We can also see this from the more explicit exact solution~\eqref{phi1-PW}, but the argument here also applies directly to the gravity case, where at large $r$ the first-order perturbation $h^1_{\mu\nu}$ can be written in a form analogous to~\eqref{phi1-BD}.

\vfill

\subsection{Matching at second order}\label{matching-2nd-order}
At second order, one could use matching to find approximations to the functions $F^2_{\ell m}$. That would be the goal if we wished to obtain second-order--accurate  waveform amplitudes, for example. However, my interest here lies only in confirming Eq.~\eqref{k2l>0} and determining $k_{200}(\tilde u)$. This reduces the matching procedure to evaluating the integral~\eqref{ret-klm} and comparing the result to Eqs.~\eqref{tildeR-22lm}, \eqref{tildeR-22lm=0}, and \eqref{tildeR-2200-v2}. %I first cast the integral in a more convient form in Sec.~\ref{integral}. I then evaluate it for $\ell=0$ in Sec.~\ref{integral-ell=0} and for for $\ell>0$ in Sec.~\ref{integral-ell>0}. 

%\subsubsection{Evaluation in the evolution zone}\label{integral}
Manipulating the integral~\eqref{ret-klm} is made easier by introducing a cutoff at $z=\tilde T:=T/\e^{n+1}$, where $T>0$ and $n>0$ are $\e$-independent constants. This is not the route taken by BD, but it is equivalent to theirs. It can be justified by splitting the integral into two pieces, $\int_r^{\tilde T} dz$ and $\int_{\tilde T}^{\infty}\! dz$, and establishing that the second integral is smaller than $O(\e^0)$. Begin by expanding around $B=0$, using $x^B=1+B\ln x+\O(B^2)$, to get
\begin{widetext}
\beq
{\rm FP}\frac{1}{K(B,2)}\hat\partial_L\left[\frac{(z-r)^{B+\ell}-(z+r)^{B+\ell}}{r}\right] = K_\ell\hat\partial_L\left[\frac{(z-r)^{\ell}\ln(z-r)-(z+r)^{\ell}\ln(z+r)}{r}\right],
\eeq
%\end{widetext}
where $K_\ell:=\frac{(-1)^\ell}{2(\ell!)}$. Next show that for large $z$ it behaves as $\sim \frac{r^\ell}{z^{\ell+1}}$, which follows from the fact that~\cite{Blanchet-Damour:86}
\beq\label{partial_L rk}
\hat\partial_L r^{k} = 0 \text{ for even integers $0\leq k<2\ell$}.
\eeq
%\beq
%{\rm FP}\frac{1}{K(B,2)}\hat\partial_L\left[\frac{(z-r)^{B+\ell}-(z+r)^{B+\ell}}{r}\right]\sim \frac{r^\ell}{z^{\ell+1}}
%\eeq
Now expand the field equations and equations of motion for large $r_p$ to obtain the approximation $S^{(-2)}_L\sim \frac{1}{r_p^5}\sim \frac{1}{(\e z)^5}$, giving us an integral of the form $\sim\e^{-5}\int_{T/\e^{n+1}}^\infty \frac{dz}{z^{6+\ell}}$. Finally, change the integration variable to $\tilde z = \e^{n+1}z$ to write the integral as $\sim \e^{n(5+\ell)}\int_{T}^\infty \frac{d\tilde z}{\tilde z^{6+\ell}}$. For $n>0$, this is negligible. Defining $\Psi_\ell:=\Box^{-1}_{\rm ret}\left(r^{-2} S^{(-2)}_L\nhat^L\right)$, we may rewrite Eq.~\eqref{ret-klm} (with $k=2$) as
%\begin{widetext}
\beq\label{ret-2lm}
\Psi_\ell = {\rm FP}\frac{K_\ell}{B} \int_r^{\tilde T} dz\, S^{(-2)}_L(t-z)\hat\partial_L\left[\frac{(z-r)^{B+\ell}-(z+r)^{B+\ell}}{r}\right]+o(\e^0),
\eeq
\end{widetext}
where ``$o(\e^p)$'' means ``goes to zero faster than $\e^p$''. I have simplified the FP operation by pulling out the pole at $B=0$ in $1/K(B,2)$.

Given that our central problem lies in the $\ell=0$ mode, and given that the analysis is significantly simpler for that mode, in the remainder of this section I examine Eq.~\eqref{ret-2lm} separately for $\ell=0$ and $\ell>0$. The reader should keep in mind that many steps in what follows rely on the fact that the integration range in Eq.~\eqref{ret-2lm} is finite. And crucially, the introduction of the cutoff relied on the fact that $S^{(-2)}_L(t-z)\to0$ as $z\to\infty$; if we had used a precisely circular orbit as our first-order source, then $S^{(-2)}_L$ would not decay in the infinite past, and the integration would have encountered the same logarithmic divergence as in Sec.~\ref{near-zone}. %Secs.~\ref{near-zone} and \ref{multiscale}, I divided the discussion of the second-order solution into $m=0$ and $m\neq0$ modes. However, because the main problem occurs in the $\ell=0$ mode, and because the evaluation of the integral~\eqref{ret-klm} is simplest for that mode, here I divide the discussion into $\ell=0$ and $\ell>0$ modes.

\subsubsection{$\ell=0$}\label{integral-ell=0}

For $\ell=0$, Eq.~\eqref{ret-2lm} simplifies to
%\begin{widetext}
\begin{align}\label{ret-200}
\Psi_0 &= \frac{Y_{00}}{2} \int_r^{\tilde T} dz\, S^{(-2)}_{200}(t-z)\frac{\ln(z-r)-\ln(z+r)}{r}\nonumber\\
				&\quad +o(\e^0),
\end{align}
where I have used Eq.~\eqref{SLnL=SlmYlm}. I split this into two integrals, $\int_r^\infty dz\, S^{(-2)}(t-z)\frac{\pm\ln(z\mp r)}{r}$, and perform a change of variables to $s=z\mp r$. Noting that the change in the upper limit has negligible effect, we find
\begin{align}\label{ret-200-v2}
\Psi_0 &= \frac{Y_{00}}{2r}\int_0^{\tilde T}\!\! ds\!\left[S^{(-2)}_{200}(u-s)-S^{(-2)}_{200}(u-s+2r)\right]\! \ln s \nonumber\\
						&\quad +\frac{Y_{00}}{2r}\int_0^{2r}\!\!ds\, S^{(-2)}_{200}(u-s+2r)\ln s+o(\e^0).
\end{align}

I now substitute the multiscale approximation of $F^1_L(u,\e)$, implying $S^{(-2)}_{200}(t)=\tilde S^{(-2)}_{200}(\e t)+\O(\e)$, where $\tilde S^{(-2)}_{200}$ is exactly as calculated in Sec.~\ref{multiscale}. One might distrust this substitution, since the integration range $\sim 1/\e^{n+1}$ is much larger than the multiscale expansion's naive domain of validity $\sim1/\e$. However, standard theorems in singular perturbation theory suggest that the expansion's domain extends to $s\lesssim 1/\e^{1+n}$ for some $n>0$; and as argued in Sec.~\ref{multiscale}, the multiscale approximation to $\varphi_1$ (and hence to $S_2$) is likely accurate on all scales. Given the substitution, I expand $\tilde S^{(-2)}_{200}$ as
\begin{align}
\tilde S^{(-2)}_{200}(\tilde u-\e s+2\e r) &= \tilde S^{(-2)}_{200}(\tilde u-\e s)+2\e r\dot{\tilde S}^{(-2)}(\tilde u-\e s)\nonumber\\
						&\quad +\O(\e^2),\label{S-expansion}
\end{align}
yielding
%where an overdot indicates a derivative with respect to the argument. The scaling with $\e$ comes from the time scale over which $S^{(-2)}$ varies, which is $\sim 1/\e$ in the evolution zone and $\gtrsim 1/\e$ in the remote past. Given this approximation, Eq.~\eqref{ret-200-v2} simplifies to
\begin{align}\label{ret-200-v3}
\Psi_0 &= -Y_{00}\int_0^{\tilde T}\! ds\,\e\dot{\tilde S}^{(-2)}_{200}(\tilde u-\e s)\ln s \nonumber\\
																&\quad +\frac{Y_{00}}{2r}\int_0^{2r}\!\!ds\, \tilde S^{(-2)}_{200}(\tilde u-\e s)\ln s +o(\e^0).
\end{align}
 
%\footnote{The validity of this is obvious in the integral $\int_0^{2r}\!\!ds$. Its validity in the integral $\int_0^\infty\! ds$ can be established by dividing the integration domain into two pieces, $\int_0^{T/\e^{1+n}} ds$ and $\int_{T/\e^{n+1}}^{\infty}\! ds$, where $n>0$ and $T>0$ are $\e$-independent constants. 
%The integral $\int_{T/\e^{n+1}}^{\infty}\! ds$ covers a range of $s$ at which $r_p(u-s)\gtrsim 1/\e^n \gg 1$, and we can use this fact to straightforwardly show that the integral is negligible. First expand the field equations and equations of motion for large $r_p$ to obtain the approximation $S^{(-2)}\sim 1/r_p^5$. Next change the integration variable to $\tilde r_p(s)=\e^n r_p(u-s)$ to see that the integral vanishes when $\e\to0$. 
%The integral $\int_0^{T/\e^{1+n}} ds$ includes a region outside the multiscale expansion's naive domain of validity $s\lesssim 1/\e$. However, from standard theorems in singular perturbation theory, we can expect the expansion's domain to extend to $s\lesssim 1/\e^{1+n}$ for some $n>0$. Hence, the only remaining step is to show that we can take $T\to\infty$---or equivalently, show that $\e\int_{T/\e^{1+n}}^{\infty} ds\, \dot{\tilde S}^{(-2)}(\tilde u-\e s)\ln s$ is negligible. The latter follows immediately after changing the integration variable to $\tilde s=\e s$.} 

After a change of integration variable to $\tilde s=\e s$, the first integral becomes
\begin{align}
\int_0^{\tilde T}\!\! ds\,\e\dot{\tilde S}^{(-2)}_{200}(\tilde u-\e s) \ln s &= \int_0^{T/\e^n} \!\! d\tilde s \, \dot{\tilde S}^{(-2)}_{200}(\tilde u-\tilde s)\ln \tilde s \nonumber\\
															&\quad + \tilde S_{200}(\tilde u)\ln\e +o(\e^0).
\end{align}
With the expansion $\tilde S^{(-2)}_{200}(\tilde u-\e s)=\tilde S^{(-2)}_{200}(\tilde u)+\O(\e)$, the second integral evaluates to 
\begin{align}
\frac{1}{2r}\int_0^{2r}\!\!ds\, \tilde S^{(-2)}_{200}(\tilde u-\e s)\ln s &= [\ln(2r)-1]\tilde S^{(-2)}_{200}(\tilde u) \nonumber\\
																		&\quad +\O(\e).
\end{align}
%Assuming $S^{(-2)}(u)$ is negligble for $u\gg1/\e$ {\bf is this valid? negligible in what sense?}, we may use 

These results combine to give us
\begin{align}
\Box^{-1}_{\rm ret}\left(r^{-2} S^{(-2)}_{200}\right) &= \left(\ln \frac{2r}{\e}-1\right)\tilde S^{(-2)}_{200}(\tilde u) \nonumber\\
					&\quad - \int_0^\infty\!\! d\tilde s \, \dot{\tilde S}^{(-2)}_{200}(\tilde u-\tilde s)\ln \tilde s \nonumber\\
					&\quad +o(\e^0).\label{R200-final}
\end{align}
Here I have changed the upper limit from $T/\e^n$ to $\infty$, which has an $o(\e^0)$ effect; the result is an integral that, at fixed $\tilde u$, is independent of both $r$ and $\e$.

Since $\varphi_{200}=\Box^{-1}_{\rm ret}\left(r^{-2} S^{(-2)}_{200}\right)+\O(1/r)$, Eq.~\eqref{R200-final} provides the leading large-$r$ behavior of the second-order monopole. It must agree with the previous expression~\eqref{tildeR-2200-v2} from the multiscale expansion, which fixes the previously unknown function $k'_{200}(\tilde u)$ to be
\begin{align}
k'_{200} &= -\tilde S^{(-2)}_{200}(\tilde u)\left(1+ \ln\frac{\e}{2}\right) \nonumber\\
				&\quad - \int_0^\infty\! \!d\tilde s \, \dot{\tilde S}^{(-2)}_{200}(\tilde u-\tilde s)\ln \tilde s.\label{k'200-final}
\end{align}

With this result, the infrared divergence is resolved. The final term in Eq.~\eqref{k'200-final} shows that the divergence was caused by neglecting hereditary effects in the wave propagation, which could not have been determined within the near-zone expansion. The first term in Eq.~\eqref{k'200-final} shows that these hereditary effects introduce $\ln\e$ terms into the field, a well-known fact in PN theory; again, this logarithm could not have been determined without knowledge of the solution outside the near zone.

In addition to resolving the infrared divergence, the calculations in this section also elucidate its cause. In Eq.~\eqref{S-expansion}, we can plainly see the failure of the multiscale expansion in the far zone at second order: an expansion in powers of $\e$ at fixed $\tilde u$ involves expanding in powers of $\e r$, restricting the range of $r$ to values $\ll 1/\e$. This occurs because the second-order field at $r$ is sourced by nonlinearities in the past history at both lesser and greater $r$, manifesting in the dependence on both $t+r$ and $t-r$ in Eq.~\eqref{ret-200-v2}.

\subsubsection{$\ell>0$ modes}\label{integral-ell>0}
Our only remaining task is to confirm that $k_{2\ell m}=0$ for all $\ell>0$.

I begin by further simplifying Eq.~\eqref{ret-2lm}, following steps in BD. As in the $\ell=0$ case, I split the integral into two, $\int_r^{\tilde T} dz\, S^{(-2)}_L(t-z)\hat\partial_L\frac{\pm(z\mp r)^{B+\ell}}{r}$. I further split the second integral into  $\int_{-r}^{\tilde T} dz-\int_{-r}^r dz$. The derivatives can then be moved outside the integrals $\int_{\pm r}^{\tilde T}dz$, since derivatives of the lower limits force evaluation at $z=\pm r$. After a change of variables to $s=z\mp r$, I combine the two integrals of the form $\int_0^{\tilde T}ds$ and move the derivatives back inside. The result is 
%\begin{widetext}
\begin{align}
\Psi_\ell &= {\rm FP}\frac{K_\ell}{B}\int_{-r}^{r} dz\,S^{(-2)}_L(t-z)\hat\partial_L\frac{(z+r)^{B+\ell}}{r}
			\nonumber\\ &\quad +X_\ell+o(\e^0),\label{Psi_ell>0}
\end{align}
where 
\beq\label{X}
X_\ell:=K_\ell\int_0^{\tilde T}\!\!\! ds\ s^{\ell}\ln s\, \hat\partial_L\frac{S^{(-2)}_L(u-s)-S^{(-2)}_L(v-s)}{r},
\eeq
and $v=t+r$. In arriving at this, I have allowed myself to omit integrals of the form $\int_{\tilde T}^{\tilde T+r}ds$ and $\int_{\tilde T-r}^{\tilde T}ds$; although they are not necessarily negligible, they will be cancelled by $T$-dependent terms in Eq.~\eqref{X}.

I first focus on $X_\ell$ and show it to be vanishingly small (up to the $T$-dependent terms mentioned above); this is exactly analogous to BD's PN analysis, where it was found that such tail terms behave as $1/c^{\ell}$. 

Using $\hat\partial_L f(r) = r^\ell\nhat^L\left(\frac{1}{r}\partial_r\right)^\ell f(r)$, Eq.~\eqref{SLnL=SlmYlm}, and $S^{(-2)}_{\ell m}(t)=\tilde S^{(-2)}_{2\ell m}(\e t)e^{-im\tilde\phi_p(\e t)}+\O(\e)$, I split $X_{\ell}$ into modes $X_{\ell m}$. For the $m=0$ modes, $X_{\ell 0}=o(\e^0)$ can be verified by roughly following BD. The simplest procedure is to change variables to $\tilde s=\e s$, expand both $\tilde S^{(-2)}_{2\ell m}(\e u-\tilde s)$ and $\tilde S^{(-2)}_{2\ell m}(\e v-\tilde s)$ around $\tilde S^{(-2)}_{2\ell m}(\e t-\tilde s)$, apply the radial derivatives, and then re-expand around $\tilde S^{(-2)}_{2\ell m}(\tilde u-\tilde s)$. Appealing to Eq.~\eqref{partial_L rk} then establishes the desired result.

For the $m\neq0$ modes, the presence of the fast time $\tilde \varphi_p$ necessitates a more involved analysis. I begin by writing $X_{\ell m}$ in a form that makes manifest its rapidly oscillating integrand,
\beq\label{X-schematic}
X_{\ell m}:=\frac{K_\ell}{\e^{\ell+1}}\int_0^{T/\e^n}\!\!\! d\tilde s\ G_{\ell m}(\tilde s)e^{-im\bar\phi_p(\tilde u -\tilde s)/\e},
\eeq
where I have introduced $\bar\phi_p:=\e\tilde\phi_p\sim\e^0$ to pull out the factor of $1/\e$ in Eq.~\eqref{phip-expansion}. The function $G_{\ell m}$ can be written as $G_{\ell m}=\tilde s^\ell \ln\frac{\tilde s}{\e}\,\tilde S^{(-2)}_{2\ell m}(\tilde u -\tilde s)r^\ell\left(\frac{1}{r}\partial_r\right)^\ell\frac{-e^{-2im\Omega r}}{r}+o(\e)$ (after expanding functions of $\e v$ around their values at $\e u$). Taking advantage of the rapid oscillations, I integrate by parts $\ell$ times to obtain
\begin{align}
X_{\ell m} &= \frac{K_\ell}{\e^{\ell+1}}\sum_{k=1}^\ell (-1)^{k} \e^k V_m(0)\frac{d^{k-1}G_{\ell m}}{d\tilde s^{k-1}}(0)e^{-im\bar\phi_p(\tilde u)/\e}\nonumber\\
						&\quad+o(\e^0)\label{X-integrated}
\end{align}
where $V_m(\tilde s) = \sum_{j\geq0}\e^j V_{mj}(\tilde s)$, $V_{m0}(\tilde s)=\frac{-1}{im\Omega(\tilde u-\tilde s)}$, and $V_{m,j+1}=\frac{\dot V_{mj}}{im\Omega(\tilde u-\tilde s)}$. In Eq.~\eqref{X-integrated} I have again omitted potentially nonnegligible $T$-dependent terms, which cancel those mentioned below Eq.~\eqref{X}. Because of the overall factor of $\tilde s^\ell$ in $G_{\ell m}$, $\frac{d^{k-1}G_{\ell m}}{d\tilde s^{k-1}}(0)=0$ unless $k\geq\ell+1$. Therefore, the sum in Eq.~\eqref{X-integrated} vanishes, leaving us with $X_{\ell m}=o(\e)$.

Hence, $\Psi_\ell$ reduces to the integral $\int_{-r}^{r} dz$ in Eq.~\eqref{Psi_ell>0}. To evaluate it, I again decompose it into $m$ modes, this time using~\cite{Blanchet-Damour:86}
\begin{align}
\hat\partial_L\frac{(z+r)^{B+\ell}}{r} &= \sum_{j=0}^\ell a_{\ell j}\frac{\nhat_L}{r^{j+1}}\frac{d^{\ell-j}}{dz^{\ell-j}}(z+r)^{B+\ell}\\
 &= \sum_{j=0}^\ell a_{\ell j}\frac{\nhat_L}{r^{j+1}}\frac{(B+\ell)!}{(B+j)!}(z+r)^{B+j},
\end{align}
where $a_{\ell j}:=\frac{(\ell+j)!}{(-2)^jj!(\ell-j)!}$. I also use $\tilde S^{(-2)}_{2\ell m}(\e t-\e z)=\tilde S^{(-2)}_{2\ell m}(\tilde u)+\O(\e)$ inside the integrand. With $\Psi_{\ell m}$ defined via $\Psi_{\ell m}Y_{\ell m}={\rm FP}\Box^{-1}_{\rm ret}(r^{B-2}S^{(-2)}_{2\ell m}Y_{\ell m})$, these steps bring Eq.~\eqref{Psi_ell>0} to the form
\begin{align}
\Psi_{\ell m} &= {\rm FP}\frac{K_\ell}{B}\sum_{j=0}^\ell \frac{a_{\ell j}}{r^{j+1}}\frac{(B+\ell)!}{(B+j)!}
			\tilde S^{(-2)}_{2\ell m}(\tilde u)\nonumber\\
			&\quad \times\int_{-r}^{r} dz\,e^{-im\tilde\phi_p(\e t-\e z)}(z+r)^{B+j}+o(\e^0).\label{Psi_ell>0v2}
\end{align}

For $m=0$, the integral is easily evaluated. After expanding around $B=0$ and taking the finite part, it becomes
\begin{align}
\Psi_{\ell 0} &= \tilde S^{(-2)}_{2\ell 0}(\tilde u)\sum_{j=0}^\ell\frac{(-1)^{\ell-j}(\ell+j)!}{(j!)^2(\ell-j)!(j+1)}\nonumber\\
			&\quad \times[\ln (2r)+\psi(\ell+1)-\psi(j+2)]+o(\e^0)\nonumber\\
			&= -\frac{\tilde S^{(-2)}_{2\ell 0}(\tilde u)}{\ell(\ell+1)} +o(\e^0),\label{Psi_ell,m=0}
\end{align} 
where in the first line, $\psi$ is the digamma function.

For $m\neq0$, I write $e^{-im\tilde\phi_p(\e t-\e z)}=e^{-im[\tilde\phi_p(\tilde u)+im\tilde\Omega_0(\tilde u)r]}e^{im\tilde\Omega_0(\tilde u)z}+\O(\e)$. The integral can then be exactly evaluated and expanded at large $r$ to yield ${\rm FP}\frac{1}{B}\int_{-r}^{r} dz\,e^{im\tilde\Omega_0(\tilde u)z}(z+r)^{B+j}=e^{im\tilde\Omega_0(\tilde u)r}p(r)$, where $p(r)$ is a power series in $1/r$. Explicitly,
\begin{align}
\Psi_{\ell m} &= \frac{\ln(r) \tilde S^{(-2)}_{2\ell m}(\tilde u)e^{-im\tilde\phi_p(\tilde u)}}{2im\tilde \Omega_0 r}\sum_{j=0}^\ell\frac{(-1)^{\ell-j}(\ell+j)!}{(j!)^2(\ell-j)!}\nonumber\\
			&\quad+O(r^{-1})+o(\e^0)\nonumber\\
			&= \frac{\ln(r) \tilde S^{(-2)}_{2\ell m}(\tilde u)e^{-im\tilde\phi_p(\tilde u)}}{2im\tilde \Omega_0 r}  +O(r^{-1}) \nonumber\\
			&\quad+o(\e^0),\label{Psi_ell,mneq0}
\end{align} 
where the $O(r^{-1})$ remainder has the form ``constant$/r$''$+\O(r^{-2}\ln r)$.

We now obtain our final result. Comparison of Eq.~\eqref{Psi_ell,m=0} to \eqref{tildeR-22lm=0} confirms that $k'_{2\ell0}=0$, since no terms of the form $r^\ell$ appear in Eq.~\eqref{Psi_ell,m=0}. Similarly, comparison of Eq.~\eqref{Psi_ell,mneq0} to \eqref{tildeR-22lm} confirms that $k'_{2\ell m}=0$ for $m\neq0$, since no terms of the form ``oscillation/$r$'' appear in Eq.~\eqref{Psi_ell,mneq0}.

%\end{widetext}

%%%%%%%%%%%%%%%%%%%%%%%%%%%%%%%%%%%%%%%%%%%%%%%%%%%%%%%
\section{Snapshot solutions and conservative dynamics}\label{snapshots}
With the multiscale expansion and matching procedure, I focused on building solutions that remain accurate on large temporal and spatial scales. But the starting point of this paper in Sec.~\ref{near-zone} was the consideration of behavior on the orbital timescale in the near zone, and oftentimes that is the only behavior we are interested in---a ``snapshot'' of the system. In this section, I show how, with a puncture at infinity, we can obtain such snapshot solutions using the near-zone expansion in Sec.~\ref{near-zone}.

We are particularly interested in these snapshots as a way of defining and computing the conservative dynamics of the system. As stressed in Ref.~\cite{Pound:14c}, beyond linear order in perturbation theory, there is no unique split into conservative and dissipative effects. Here I adopt the definition used in Ref.~\cite{Pound:14c}, which is equivalent to working with retarded fields but simply setting the dissipative forces $f^t_{\rm self}$ and $f^\phi_{\rm self}$ to zero in the equation of motion for the quasicircular orbit. From a global perspective, this would result in an eternally circular orbit, leading, as we now realise, to a divergent second-order solution. However, within the context of a snapshot, we do not encounter that problem.

To see how this plays out, begin with the multiscale solution, write $\tilde u=\e(t-r)$, and then expand for small $\e$. This yields a Gralla-Wald expansion, $\e\varphi_{1}+\e^2\varphi_2=\e\GW{\varphi}_1+\e^2\GW{\varphi}_2+o(\e^2\ln\e)$, that is accurate in the near zone on the orbital timescale around $t=0$. Without loss of generality, I single out $t=0$ as the time at which Eq.~\eqref{r1(0)} holds, such that the Gralla-Wald expansion is an expansion at fixed frequency $\GW{\Omega}_0 = \tilde\Omega_0(0)$, allowing me to set $\tilde\Omega_1(0)$ terms to zero. The first-order field is then given by
\beq
\GW{\varphi}_{1\ell m} = \tilde R_{1\ell m}(0,r)e^{-im\GW{\Omega}_0 u}.
\eeq
Following Sec.~\ref{near-zone}, the second-order field cleanly splits into two terms, $\GW{\varphi}_2=\GW{\psi}+\delta\varphi_1$, given by
\begin{align}
\GW{\psi}_{\ell m} &= \tilde R^{\tilde\psi}_{2\ell m}(0,r)e^{-im\GW{\Omega}_0u},\\
\delta\varphi_{1\ell m} &= \left[\dot{\tilde R}_{1\ell m}(0,r)u
													-\tfrac{1}{2}im\dot{\tilde\Omega}_0(0)\tilde R_{1\ell m}(0,r)u^2\right.\nonumber\\
												&\quad \left.+\tilde R^{\widetilde{\delta\varphi}_1}_{2\ell m}(0,r)\right]e^{-im\GW{\Omega}_0u},\label{deltavphi}
\end{align}
where the fields $\tilde R^{\tilde\psi}_{2\ell m}$ and $\tilde R^{\widetilde{\delta\varphi}_1}_{2\ell m}$ are as described in Sec.~\ref{multiscale-2nd-order-field}. Specifically, $\tilde R^{\widetilde{\delta\varphi}_1}_{2\ell m}(0,r)$ is sourced by $-4\pi\tilde\varrho_{2\ell m}(0,r)+2(\partial_r+1/r)\dot{\tilde R}_{1\ell m}(0,r)$, with $\tilde\varrho_{2\ell m}$ given by Eq.~\eqref{varrho2}.

Now, the conservative sector of the solution, as defined above, is obtained by setting $\dot{\tilde r}_0(0)=0$. This eliminates most terms in Eq.~\eqref{deltavphi}, reducing $\delta\varphi_{1\ell m}$ to a certain periodic piece $\delta\varphi^{\rm c}_{1\ell m}e^{-im\GW{\Omega}_0t}$ sourced by $-4\pi\delta\rho_{\ell m}^{\rm c}e^{-im\GW{\Omega}_0t}$, where
\begin{align}
\delta\rho^{\rm c}_{\ell m}(r) &= -\frac{N_{\ell m}}{\GW{U}_0r^2}\left[\GW{r}^c_1\delta'(r\!-\!\GW{r}_0\!)
																			+\frac{\GW{U}^{\rm c}_1}{\GW{U}_0}\delta(r\!-\!\GW{r}_0)\right].
\end{align}
Here $\GW{r}^{\rm c}_1=\tilde r_1(0)=r_{10}$ [referring to Eq.~\eqref{scaling}], and $\GW{U}^{\rm c}_1=\tilde U_1(0)=-\GW{r}^{\rm c}_1/(2\GW{r}_0^2\GW{U}_0^3)$. The total solution in the conservative sector is then $\varphi^{\rm c}_{\ell m}=\e\GW{\varphi}_{1\ell m}+\e^2(\GW{\psi}_{\ell m}+\delta\varphi^{\rm c}_{1\ell m})$.

With the exception of the $\ell=0$ mode, this solution is identical to what we would have obtained in Sec.~\ref{near-zone} if we had set the dissipative components of the force to zero in the equation of motion~\eqref{EOM} and chosen the worldline $z^\mu$ to be a circular orbit of radius $r_p=\GW{r}_0+\e \GW{r}^{\rm c}_{1}+\O(\e^2)$ and of frequency $\Omega=\GW{\Omega}_0$. For the $\ell=0$ mode, we would also have obtained the above solution with the procedure in Sec.~\ref{near-zone} if we had combined it with the puncture at infinity described in Sec.~\ref{multiscale-2nd-order-field}. 

Concretely, the first-order solution is unaltered from Eqs.~\eqref{phi1-lmw} and \eqref{GWphi1}. In the second-order solution, because $\delta\varphi^{\rm c}_{1\ell m}$ has none of the growing terms (neither temporally nor spatially) in $\delta\varphi_{1\ell m}$, the split into $\GW{\psi}$ and $\delta\varphi_1$ is unnecessary, and we may write $\GW{\varphi}_{2\ell m}^{\rm c}=\GW{R}^{\rm c}_{2\ell m}(r)e^{-im\GW{\Omega}_0t}$. For $\ell>0$, the radial functions $\GW{R}^{\rm c}_{2\ell m}$ can be obtained from Eq.~\eqref{GWphi2} with a change in source $\GW{S}_{2\ell m}\to \GW{S}^{\rm c}_{2\ell m}$ in the integrands of $\GW{c}^\pm_{2\ell m}$, where the new source includes the contribution of $\delta\rho^{\rm c}_{\ell m}$,
\beq
\GW{S}^{\rm c}_{2\ell m} = \GW{S}_{2\ell m}-4\pi\delta\rho^{\rm c}_{\ell m}.
\eeq
For $\ell=0$, $\GW{R}^{\rm c}_{2\ell m}$ can be  obtained indirectly from Eq.~\eqref{GWphi2} via the puncture at infinity. Introduce the effective field $\GW{R}^{\rm eff}_{200}$, related to $\GW{R}^{\rm c}_{200}$ by
\beq\label{GWR-eff}
\GW{R}^{\rm c}_{200} =  \GW{R}^{\rm eff}_{200} +\GW{\varphi}^\infty + \GW{k}_{200},
\eeq
where $\GW{\varphi}^\infty$ and $\GW{k}_{200}$ are the leading terms in the re-expansion of Eqs.~\eqref{phi-infty} and \eqref{k'200-final}. Explicitly, 
\begin{align}
\GW{\varphi}^\infty &= \theta(r-\GW{r}^{\infty})\ln(r) \GW{S}^{(-2)}_{200},\label{GWphi-infty}\\
\GW{k}_{200} &= -\left(1+ \ln\frac{\e}{2}\right)\GW{S}^{(-2)}_{200} \nonumber\\
				&\quad - \int_0^\infty\! \!d\tilde s \, \dot{\tilde S}^{(-2)}_{200}(-\tilde s)\ln \tilde s,\label{GWk}
\end{align}
where $\GW{r}^{\infty}>\GW{r}_0$ is an arbitrary constant. The effective field $ \GW{R}^{\rm eff}_{200}$ can be obtained from Eq.~\eqref{GWphi2} with a change in source $\GW{S}_{2\ell m}\to \GW{S}^{\rm eff}_{2\ell m}$, where 
\beq
\GW{S}^{\rm eff}_{200} = \GW{S}_{200}-4\pi\delta\rho^{\rm c}_{00}-(\partial_r^2+2r^{-1}\partial_r)\GW{\varphi}^\infty.
\eeq
The full field $\GW{R}^{\rm c}_{200}$ can then be recovered from Eq.~\eqref{GWR-eff}.

With this method, we can construct a complete snapshot of the conservative sector of the solution in the near zone. Unfortunately, because Eq.~\eqref{GWk} contains an infinite time integral, the solution is not self-contained: we cannot determine the constant $\GW{k}_{200}$ without determining $\dot{\tilde S}^{(-2)}_{200}$ from the multiscale expansion. However, as mentioned previously, a constant shift in $\varphi$ is a gauge freedom in the model. Hence, $\GW{k}_{200}$ has no effect on the system's dynamics, and in the end it may be neglected.

%Note that $\delta\varphi^{\rm c}_{1\ell m}$ is not the only piece of $\delta\varphi_{1\ell m}$ that is constant along the orbit, and we cou

%The coefficients read 
%\begin{align}
%\GW{R}_{1\ell m} &=\tilde{R}_{1\ell m}(0,r)\\
%\GW{R}_{2\ell m} &=
%\end{align}

%%%%%%%%%%%%%%%%%%%%%%%%%%%%%%%%%%%%%%%%%%%%%%%%%%%%%%%
\section{Lessons for gravity}\label{lessons}
In this paper, I have shown how problems of large scales arise in nonlinear field theories with slowly evolving sources. More importantly, I have presented methods to overcome these problems. Although the methods were demonstrated for a simple toy problem, they should carry over to the relevant problem of a compact object orbiting a large black hole. In the remainder of this section, I discuss what changes need to be made and what work remains to be done.

%The accuracy is hence limited to a strip covering a large range of $u$ but a smaller range of $r$. However, this limitation arises only in $\varphi_2$, not in $\varphi_1$.

\subsection{Matching procedure and conservative dynamics}
In principal, second-order self-force computations have been possible for some time. Reference~\cite{Pound:14c} showed how to set up a meaningful calculation of a conservative effect on a quasicircular orbit in Schwarzschild within a Gralla-Wald expansion, Ref.~\cite{Wardell-Warburton:15} developed a method and numerical infrastructure for implementing such a calculation in the frequency domain, and Ref.~\cite{Pound-Miller:14} provided the necessary analytical input. 

However, implementation has been hampered by the presence of infrared divergences, which arise in precisely the manner described in Sec.~\ref{near-zone}. Physically, they occur because if we extend the integrals to infinity while using a circular orbit as a first-order source, the solution behaves as if the particle has \emph{always} been on that circular orbit, radiating energy at a constant rate, such that each spatial hypersurface stores infinite energy. 

Just as in the toy model, this obstacle can be overcome with the matching procedure inherited from PN theory. In the gravity case, we match an expansion around a Schwarzschild background to a PM expansion around Minkowski. Because the equations~\eqref{EFE1}--\eqref{EFE2} for the metric perturbations $h^n_{\mu\nu}$ reduce to flat-space wave equations at leading order in $1/r$, the mathematics of the matching is essentially identical to Sec.~\ref{matching}.

Once the matching is performed, we can compute conservative effects using a snapshot of the solution on the orbital timescale, just as described in Sec.~\ref{snapshots}. After decomposing into a basis of tensor harmonics, for the $\ell>0$ modes we may use the field equations~\eqref{EFE1}--\eqref{EFE2} in (the decomposed version of) their Gralla-Wald form given by Eqs.~(77) and (79) in Ref.~\cite{Pound:14c}. These equations can be solved using Eq.~(4.24) in Ref.~\cite{Wardell-Warburton:15}. For the $\ell=0$ modes, we must modify the setup in Ref.~\cite{Pound:14c} by introducing a puncture at infinity. The puncture will take the logarithmic form~\eqref{GWphi-infty}, with only cosmetic differences. 

This determines the solution up to the addition of a constant of the form~\eqref{GWk}, knowledge of which would require evaluation of an integral over the infinite past. However, there is strong reason to believe that just as in the toy model, the integral term is pure gauge; as shown in Ref.~\cite{Blanchet-Damour:88}, monopolar hereditary integrals can be removed with a gauge transformation.

Unfortunately, even after resolving the behavior at infinity, one open problem remains. Unlike in the toy model, the problem at infinity is complemented by the same problem at the horizon of the large black hole: the first-order solution behaves as if the horizon has been absorbing energy at a constant rate for all time, which will lead to a divergent second-order solution. This is now the major obstacle to implementation. To overcome it, we must perform a matching procedure at the horizon, where, unfortunately, no nonlinear equivalent of the PM expansion is ready at hand.

%\subsection{Gralla-Wald expansion and conservative dynamics}

\subsection{Multiscale expansion and wave generation}
Numerical computations of second-order self-force effects have so far focused on conservative dynamics on the orbital timescale. But ultimately, one would like to simulate complete inspirals and generate waveforms. Fortunately, once the matching procedures at infinity and the horizon have been performed, the multiscale expansion  in Sec.~\ref{multiscale}, as summarized in Sec.~\ref{multiscale-summary}, should apply with little change in the gravity case. 

For quasicircular inspirals in Schwarzschild, the equation of motion can be expanded  as in Sec.~\ref{multiscale-motion}. The field equations can be expanded  as in Sec.~\ref{multiscale-field} with only two changes: First, the homogeneous solutions in Eq.~\eqref{general-solution} will not vanish at first order. Instead, they will include perturbations to the background mass and spin of the large black hole, and their slow evolution will be determined from the second-order Einstein equation. Second, the choice of slow time variable must be reconsidered. In flat spacetime, the field could be written in a uniform way in terms of $u$. However, in a black hole spacetime, in addition to propagating to asymptotic infinity along outward null curves, solutions propagate to the horizon along ingoing null curves. In analogy with the behavior $e^{-im\phi_p(\e u)}$ at null infinity, we can expect the exact solution to behave as $e^{-im\phi_p(\e v)}$ near the horizon. If in that region we use a slow time coordinate other than $\e v$, such as $\e t$ or $\e u$, then we introduce large errors. Hence, we likely require a slow time coordinate that behaves as $\e u$ near future null infinity and like $\e v$ near the future horizon. This can be achieved with hyperbolic slicing, which was adapted to the frequency domain in Ref.~\cite{Zenginoglu:11} and can be easily implemented in the Lorenz gauge~\cite{Warburton:private}.

Such a scheme will generate waveforms with highly accurate phase evolutions, but the amplitudes will be limited to first-order accuracy, a fact discussed in Secs.~\ref{multiscale-summary} and \ref{matching-2nd-order}. This arises due to the small error in the phase of the first-order perturbation $h^1_{\mu\nu}$, as described in Sec.~\ref{multiscale-summary}, and a failure of the multiscale expansion of $h^2_{\mu\nu}$ in the far zone, as described in Sec.~\ref{matching-2nd-order}. The second of these problem can likely be overcome by reorienting the matching procedure to focus on large-$r$ behavior. 

Of course, beyond this limitation in accuracy, the multiscale expansion considered here is also limited to quasicircular inspirals. For more general inspirals, a more general expansion must be used. In Ref.~\cite{Hinderer-Flanagan:08,Flanagan-Hinderer:12}, we already have a multiscale expansion of the equation of motion for generic inspirals in Kerr. Hence, the remaining work to be done is to combine this with a multiscale expansion of the field equations. If we restrict our attention to inspirals in Schwarzschild, the expansion of the field equations should be a straightforward extension from the quasicircular case. The expansion in Kerr will encounter greater challenges. First, orbital resonances will introduce a new timescale into the evolution~\cite{Flanagan-Hinderer:12}. Second, the gauges that are convenient for frequency-domain computations in Kerr~\cite{Shah-etal:12,Pound-Merlin-Barack:14,vandeMeent-Shah:15} are not known to extend to second order. Ideally, the second problem should be bypassed by reformulating the multiscale expansion to avoid involving the complete (gauge-dependent) second-order metric perturbation. This could be done by expressing the second-order dissipative force in terms of curvature scalars, which could be computed from the second-order Teukolsky equation~\cite{Campanelli-Lousto:98}.

%\subsection{Practical implementation of matching procedure: a puncture at infinity}

\acknowledgments
I thank Leor Barack and Eric Poisson for careful readings of this paper and helpful discussions, Soichiro Isoyama for fruitful conversations, and Chad Galley for an enlightening comment. I especially thank Takahiro Tanaka for instrumental suggestions and Alex Le Tiec for providing a reading list of post-Newtonian and post-Minkowskian literature. This work was supported by the European Research Council under the European Union’s Seventh Framework Programme (FP7/2007-2013)/ERC Grant No. 304978.

\appendix

%\section{Evolution of the deviation vector $\GW{z}^\mu_1$}

\section{Coupling of derivatives of spherical harmonics}\label{coupling formula}
In this appendix I derive the coupling formula given schematically in Eq.~\eqref{S2lm-compact} and explicitly in Eq.~\eqref{S2lm} below, along with its analogue in the multiscale expansion, beginning in both cases from the general formula in the time domain. From the general formula I also obtain the formula in the multiscale expansion. The method is based on converting all harmonics and their derivatives to spin-weighted harmonics ${}_sY_{\ell m}$ and then utilizing a formula for the integral of three ${}_sY_{\ell m}$'s.  Section~\ref{sYlm} contains the relevant identities for ${}_sY_{\ell m}$'s, and Sec.~\ref{coupling derivation} presents the derivation.

This same method was also used to derive a formula for the coupling of tensor-harmonic modes in Eq.~\eqref{second-order_Ricci} in the gravity case, reported but not given explicitly in Ref.~\cite{Pound:14c}; that formula and its derivation will be presented elsewhere. 
 
%\cite{Blanchet-Damour:86,Damour-Iyer:91}

\subsection{Conventions and identities for spin-weighted harmonics}\label{sYlm}
Spin-weighted harmonics are constructed from a complex basis on $S^2$. Let $\Omega_{AB}={\rm diag}(1,\sin^2\theta)$ be the metric on the sphere. I define 
\begin{equation}
m^A\equiv \left(1,\frac{i}{\sin\theta}\right),
\end{equation}
which is proportional to $e^A_\theta+ie^A_\phi$, with $e^A_\theta$ and $e^A_\phi$ being unit vectors (with respect to $\Omega_{AB}$) in the $\theta$ and $\phi$ directions. This vector has the useful properties
\begin{equation}\label{m identities}
\begin{array}{lll}
m^Am_A=0, &\quad m^Am_A^*=2,\\
m^B D_Bm^A =  m^A \cot\theta, &\quad m^{B*} D_Bm^A = -m^A\cot\theta, \\
\epsilon_{AB}m^B = im_A, &\quad 
\end{array}
\end{equation}
and
\beq
\Omega^{AB}=\frac{1}{2}\left(m^Am^{B*}+m^{A*}m^B\right),
\eeq
where indices are raised and lowered with $\Omega_{AB}$, and $\epsilon_{AB}$ is the Levi-Civita tensor on the sphere. My definition of $m^A$ differs by a factor of $\sqrt 2$ relative to that originally defined by Newman and Penrose~\cite{Newman-Penrose:66}, and it is normalized on the unit sphere rather than a sphere of radius $r$. In terms of $m^A$, I define the action of the derivative operators $\eth$ and $\bar\eth$ on a scalar of spin-weight $s$ as
\begin{align}
\eth v &= (m^AD_A-s\cot\theta)v,\label{eth}\\
\bar{\eth} v &= (m^{A*}D_A+s\cot\theta)v.\label{ethbar}
\end{align}
My definitions here differ by an overall minus sign relative to those of Newman and Penrose. A quantity $v$ has spin-weight $s$ if it transforms as $v\to e^{is\varphi}v$ under the complex rotation $m^A\to e^{i\varphi}m^A$. $\eth v$ raises the spin weight by 1, $\bar\eth$ lowers it by 1.

The spherical harmonics of spin-weight $s$ are defined as
\begin{equation}
{}_sY^{\ell m} \equiv \sqrt{\frac{(\ell-|s|)!}{(\ell+|s|)!}}
	\begin{cases}
    (-1)^s\eth^sY^{\ell m}, & 0\leq s\leq\ell,\\
    \bar{\eth}^{|s|} Y^{\ell m}, & -\ell\leq s\leq0.
  \end{cases}
\end{equation}
They are proportional to certain rotation matrix elements~\cite{Goldberg-etal:67}:
\begin{equation}
{}_sY^{\ell m} = \sqrt{\frac{2\ell+1}{4\pi}}D^\ell_{-sm}(\phi,\theta,0),\label{YtoD}
\end{equation}
where $D^\ell_{m' m}(\alpha,\beta,\gamma)$ is a Wigner D-matrix. 

Deriving the coupling formula~\eqref{S2lm} requires two ingredients: a transformation from $D_AY_{\ell m}$ to ${}_{\pm 1}Y_{\ell m}$, and a formula for an integral of three spin-weighted harmonics.

The desired expression for $D_AY_{\ell m}$ is established using $D_AY^{\ell m}=\frac{1}{2}(m_Am^{B*}+m^*_Am^B)D_BY^{\ell m}$ together with Eqs.~\eqref{eth} and \eqref{ethbar}, leading to
\begin{equation}\label{DYtosY}
D_A Y^{\ell m} = \frac{1}{2}\sqrt{\ell(\ell+1)}\left({}_{-1}Y^{\ell m}m_A-{}_1Y^{\ell m}m^*_A\right).
\end{equation}

The desired integral is
\begin{equation}\label{Cdef}
C^{\ell m s}_{\ell'm's'\ell''m''s''}:=\oint {}_{s}Y^{\ell m*}{}_{s'}Y^{\ell' m'}{}_{s''}Y^{\ell'' m''}d\Omega. 
\end{equation}
In the case that $s=s'+s''$, it can be evaluated by applying a formula for the integral of three D-matrices (see, e.g., Sec. 30B of Ref.~\cite{Hecht:00}), leading to
\begin{widetext}
\begin{align}\label{coupling}
C^{\ell m s}_{\ell'm's'\ell''m''s''} = (-1)^{m+s}\sqrt{\frac{(2\ell+1)(2\ell'+1)(2\ell''+1)}{4\pi}}
					\begin{pmatrix}\ell & \ell' & \ell'' \\ s & -s' & -s''\end{pmatrix}
					\begin{pmatrix}\ell & \ell' & \ell'' \\ -m & m' & m''\end{pmatrix},
\end{align}
where the arrays are $3j$ symbols. If $s=s'=s''=0$, this reduces to the standard formula for the integral of three ordinary spherical harmonics.

\subsection{Derivation of the coupling formula}\label{coupling derivation}
To decompose the source, I first write it as
\beq
S_2 = t^{\alpha\beta}\nabla_{\alpha}\varphi_1\nabla_\beta \varphi_1 = (\partial_t \varphi_1)^2+ (\partial_r \varphi_1)^2 + \frac{1}{r^2}\Omega^{AB}\partial_A\varphi_1\partial_B\varphi_1.\label{S2 pieces}
\eeq
(Here I neglect the contribution of $\Box\varphi_2^\P$, which, as discussed above Eq.~\eqref{S2lm-compact}, is not relevant for the analysis in this paper.)

The terms involving $t$ and $r$ derivatives can be expressed as a sum of spherical harmonics by substituting $\varphi_1=\sum_{\ell m}\varphi_{1\ell m}(t,r)Y_{\ell m}$ and then integrating against $Y^*_{\ell m}$. The result is
\beq
(\partial_t \varphi_1)^2 + (\partial_r \varphi_1)^2 = \sum_{\ell m}\sum_{\ell' m'}\sum_{\ell'' m''} C^{\ell m 0}_{\ell'm'0\ell''m''0} (\partial_t \varphi_{1\ell'm'}\partial_t\varphi_{1\ell''m''} + \partial_r \varphi_{1\ell'm'}\partial_r \varphi_{1\ell''m''})Y_{\ell m},
\eeq
where the coupling coefficients $C^{\ell m s}_{\ell'm's'\ell''m''s''}$ are defined in Eq.~\eqref{coupling}. The sums are restricted by the fact that $C^{\ell m s}_{\ell'm's'\ell''m''s''}$ enforces (i) $m=m'+m''$ and (ii) the triangle inequality $|\ell'-\ell''|\leq \ell\leq\ell'+\ell''$. Restriction (i) can be used to eliminate the sum over $m''$ by making the replacement $m''=m-m'$.

The term involving angular derivatives in Eq.~\eqref{S2 pieces} can be decomposed in the same way by converting the derivatives using Eq.~\eqref{DYtosY}, using Eq.~\eqref{m identities}, and then appealing to Eq.~\eqref{coupling}. The result is 
\beq
\Omega^{AB}\partial_A\varphi_1\partial_B\varphi_1 = -\frac{1}{2}\sum_{\ell m}\sum_{\ell' m'}\sum_{\ell'' m''} (C^{\ell m 0}_{\ell'm'-1\ell''m''1}+C^{\ell m 0}_{\ell'm'1\ell''m''-1}) \sqrt{\ell'(\ell'+1)\ell''(\ell''+1)}\varphi_{1\ell'm'}\varphi_{1\ell''m''}Y_{\ell m}.
\eeq
As above, these sums are restricted by $m=m'+m''$ and by the triangle inequality.

Putting these results together, we find that the coefficients in $S_2=\sum_{\ell m}S_{2\ell m}(t,r)Y_{\ell m}$ are
\begin{align}\label{S2lm-general}
S_{2\ell m} &=  \sum_{\ell'm'}\sum_{\ell''m''}\bigg[C^{\ell m 0}_{\ell'm'0\ell''m''0} (\partial_t\varphi_{1\ell'm'}\partial_t\varphi_{1\ell''m''} + \partial_r \varphi_{1\ell'm'}\partial_r \varphi_{1\ell''m''})\nonumber\\
 &\quad-\frac{1}{2r^2} (C^{\ell m 0}_{\ell'm'-1\ell''m''1}+C^{\ell m 0}_{\ell'm'1\ell''m''-1}) \sqrt{\ell'(\ell'+1)\ell''(\ell''+1)}\varphi_{1\ell'm'}\varphi_{1\ell''m''}\bigg].
\end{align}

In the Gralla-Wald expansion, with $\GW{\varphi}_1=\sum_{\ell m}\GW{R}_{1\ell m}(r)e^{-im\GW{\Omega}t}Y_{\ell m}$, this becomes $\GW{S}_2 =\sum_{\ell m}\GW{S}_{2\ell m}(r)e^{-im\GW{\Omega}_0 t}Y_{\ell m}$, where
\begin{align}\label{S2lm}
\GW{S}_{2\ell m} &=  \sum_{\ell'm'}\sum_{\ell''m''}\bigg[C^{\ell m 0}_{\ell'm'0\ell''m''0} (-m'm''\GW{\Omega}_0^2 \GW{R}_{1\ell'm'}\GW{R}_{1\ell''m''} + \partial_r \GW{R}_{1\ell'm'}\partial_r \GW{R}_{1\ell''m''})\nonumber\\
 &\quad-\frac{1}{2r^2} (C^{\ell m 0}_{\ell'm'-1\ell''m''1}+C^{\ell m 0}_{\ell'm'1\ell''m''-1}) \sqrt{\ell'(\ell'+1)\ell''(\ell''+1)}\GW{R}_{1\ell'm'}\GW{R}_{1\ell''m''}\bigg].
\end{align}
This is the source decomposition~\eqref{S2-decomposition}. Note that the restriction on the $m$'s allowed me to write the exponential factor as $e^{-im\GW{\Omega}_0 t}$.

In the multiscale expansion, with $\tilde\varphi_1=\sum_{\ell m}\tilde R_{1\ell m}(\tilde u,r)e^{-im\tilde\phi_p(\tilde u)}Y_{\ell m}$,  Eq.~\eqref{S2lm-general} becomes $\tilde{S}_{2}=\sum_{\ell m}\tilde{S}_{2\ell m}e^{-im\tilde\phi_p(\tilde u)}Y_{\ell m}$, where
\begin{align}\label{calS2lm}
\tilde{S}_{2\ell m} &=  \sum_{\ell'm'}\sum_{\ell''m''}\bigg[C^{\ell m 0}_{\ell'm'0\ell''m''0} \left(-2m'm''\tilde\Omega_0^2 \tilde{R}_{1\ell'm'}\tilde{R}_{1\ell''m''} + im'\tilde\Omega_0 \tilde{R}_{1\ell'm'}\partial_r \tilde{R}_{1\ell''m''}+im''\tilde\Omega_0\partial_r \tilde{R}_{1\ell'm'} \tilde{R}_{1\ell''m''}\right.\nonumber\\
 &\quad \left.+\partial_r \tilde{R}_{1\ell'm'}\partial_r \tilde{R}_{1\ell''m''}\right) -\frac{1}{2r^2} (C^{\ell m 0}_{\ell'm'-1\ell''m''1}+C^{\ell m 0}_{\ell'm'1\ell''m''-1}) \sqrt{\ell'(\ell'+1)\ell''(\ell''+1)}\tilde{R}_{1\ell'm'}\tilde{R}_{1\ell''m''}\bigg]\!.
\end{align}
This is the source term appearing in Eq.~\eqref{tildecalS2lm}. 

\end{widetext}

\bibliography{../bibfile}

%\bibliography{bibfile}

\end{document}